\documentclass[review]{elsarticle}

\usepackage{graphicx}% Include figure files
\usepackage{dcolumn}% Align table columns on decimal point
\usepackage{bm}% bold math
\usepackage{adjustbox}
\usepackage{tikz}
\usetikzlibrary{shapes.geometric, arrows}
\usepackage{url}
\usepackage{xcolor}
\usepackage{amsmath}
\usepackage{mathrsfs}
\usepackage{longtable}
\usepackage[normalem]{ulem}

\usepackage{lineno,hyperref}
\modulolinenumbers[5]
\begin{document}

%\preprint{APS/123-QED}

\title{Advanced storage ring lattice options based on hybrid six-bend achromat for 
	Stanford Synchrotron Radiation Lightsource upgrade}% Force line breaks with \\

\author{Pantaleo Raimondi}
\ead{raimondi@slac.stanford.edu} 
\author{Xiaobiao Huang}
\ead{xiahuang@slac.stanford.edu} 
\author{Jaehyun Kim}
\author{James Safranek}
\author{Tom Rabedeau}
\address{SLAC National Accelerator Laboratory, 2575 Sand Hill Road, Menlo Park, CA 94025}

\date{\today}% It is always \today, today,

\begin{abstract}
Three storage ring lattices have been designed as options for a future upgrade of the 
Stanford synchrotron radiation lightsource (SSRL). 
The three options differ in circumference and targeted future site, with one to be built in the 
tunnel of the present SPEAR3 ring, one as a green field ring on the SLAC campus, and 
the third in the tunnel of the decommissioned PEP-II ring. 
The lattices are based on the newly proposed hybrid 6-bend achromat (H6BA) lattice cells, which 
is ideal for pushing the photon beam brightness while achieving excellent nonlinear dynamics 
performance. The transparent matching conditions are enforced to minimize the negative impact of the 
loss of periodicity due to insertion of various long straight sections. 
Numerical optimization is performed to further improve the nonlinear dynamics. 
In addition to reaching very low emittances, the lattices can accommodate traditional off-axis 
injection and achieve beam lifetimes similar to or exceeding that of typical third generation rings. 
\end{abstract}

\maketitle
%\linenumbers
%\tableofcontents

\section{Introduction}
Stanford Synchrotron Radiation Lightsource (SSRL) is one of the earliest synchrotron radiation 
facilities. It currently operates Stanford Positron-Electron Asymmetric Ring-III (SPEAR3)~\cite{HettelEPAC04}
to serve over 2000 photon beam users annually. 
SPEAR3 was built on the footprint of its predecessor, SPEAR, with a circumference of 234~m. 
%It has a horizontal emittance of 7~nm.  
While SPEAR3 provides excellent photon beam performance consistent with a typical 
third-generation light source today, its relative competitiveness will decrease in the future 
as many other storage ring based light sources undergo upgrades to implement 
multi-bend achromat (MBA) lattices~\cite{EinfeldPAC95,LeemanMAXIV,ESRFHMBA,APSU41pm,ALSU,HEPS,PETRAIV,ELETTRA2,DIAMOND2, SOLEILU2}, which can increase the photon beam brightness by 
two to three orders of magnitude. 

It is imperative for SSRL to seek an upgrade path toward a high performance storage ring 
based facility which is on par with or exceeds the performance of similar facilities elsewhere. 
A lattice design effort has been undertaken to investigate various upgrade options. 
Storage ring lattices of three siting options are considered: (1) a lattice in the existing 
SPEAR3 tunnel to replace the current machine; (2) a green-field new ring located elsewhere on the SLAC campus; 
(3) a lattice to fit the PEP tunnel. 
The hybrid six-bend achromat (H6BA) lattice cell, recently proposed for diffraction limited 
storage rings~\cite{Pantaleo2023PRAB}, is used as the foundation of the lattice options. 
In the lattice design study, we assumed only mature technology (e.g., magnet, vacuum) that has successfully been demonstrated, for example, at ESRF-EBS~\cite{ESRFEBS}. 
While the designs aim at maximizing the photon beam brightness, off-axis injection and long 
beam lifetime are included as design goals in order to reduce the requirements on injector. 
The resulting lattice solutions have  excellent linear and nonlinear lattice performances. 

As was discussed in Ref.~\cite{Pantaleo2023PRAB}, modified from the ESRF-EBS hybrid 7-bend 
achromat (H7BA) cell~\cite{ESRFEBS,EBScomm}, the H6BA lattice cell leads to ideal optics properties, 
such as beta functions at the insertion devices matched to the photon beam, 
small natural emittance, and very large dynamic aperture (DA) and momentum acceptance (MA). 
When applied to the various SSRL upgrade scenarios, the H6BA cell knobs are optimized. 
Long injection straight sections with high beta functions are incorporated in the 
lattices. Techniques discussed in Ref.~\cite{Pantaleo2023PRAB} to match the long straight 
sections to satisfy the transparent conditions are employed. 
The combination of the optimized H6BA cells and the use of transparent long injection straight 
produces large DA and MA, which enables off-axis injection and 
Touschek lifetimes comparable or better than typical third generation light sources. 

Additionally, numeric optimization of the nonlinear beam dynamics with multi-objective, evolutionary 
algorithms is performed for the full ring lattices~\cite{yang2008global,borland2010direct,HUANG201448}. 
The optimization further improves the DA and MA and eliminates the need to use high order nonlinear
magnets in one of the cases. 
 
In this paper we report the design considerations, constraints, and optimizations for the three options and showcase the linear and nonlinear lattice performances. 
Section II, III, and IV discuss the three lattice options, respectively, 
while section V compares their photon brightness performance.  A table (Table~\ref{tab:RingPara}) is included to compare selected lattice parameters. The conclusion is given 
in Section VI.

\section{SSRLUP: The 234-m lattice}
\subsection{Linear lattice}
An upgrade option which could re-utilize some of the existing SSRL infrastructure involves building a storage ring on the footprint 
of the existing SPEAR3 ring. 
The SPEAR3 ring consists of 18 double-bend achromat (DBA) cells on a race-track shape, 14 of which 
are standard cells and 4 are matching cells. The bending angle in a matching cell is 
3/4 of that of the standard cell. 
To retain the same overall geometry and insertion device (ID) beamline source points, the upgrade lattice should have 
the same number of cells, the same cell lengths, and the same bending angles for each cell. 
By replacing the DBA cells with H6BA cells and properly matching the long straight sections, 
a full ring lattice was obtained. It is referred to as the SSRLUP lattice. 

\begin{figure}[t]
\includegraphics[width=0.8\textwidth]{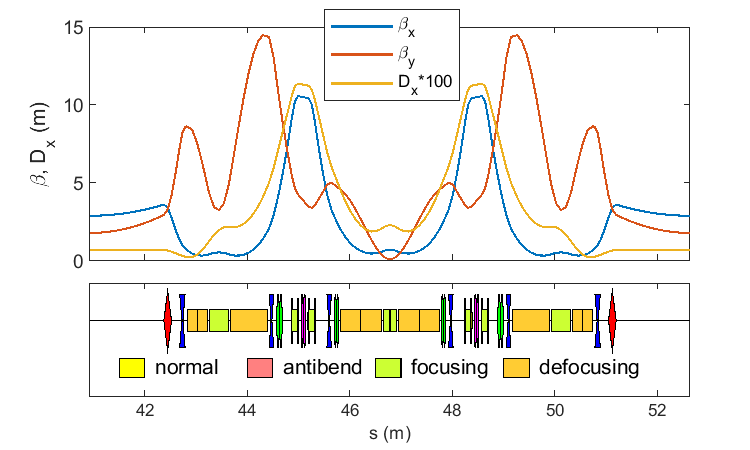}
\caption{The beta and dispersion functions for a standard cell in the SSRLUP lattice. 
The cell length is 11.7~m. 
Horizontal and vertical betatron phase advances are $\psi_x=1.7917\times2\pi$ and $\psi_y = 0.8426\times2\pi$, respectively. 
In the lattice cell layout, quadrupoles are marked by red (focusing) and blue 
(defocusing) lens shapes, respectively. Bending magnets are marked by squares with different colors to indicate if they are anti-bends or transverse gradients (same color code for figures below). 
%Normal bending magnets (positive bending angle, no transverse gradient) are marked by yellow squares, while bends with focusing gradient are in greenish  squares (as shown); bends with defocusing gradients are in brownish squares (not shown). 
\label{figSSRLcellOptics}}
\end{figure}

Figure~\ref{figSSRLcellOptics} shows the linear optics of a standard H6BA cell for the 
SSRLUP lattice. Lattice evaluation and tracking simulations in this study are done with the modeling and simulation code Accelerator Toolbox~\cite{TerebiloAT01}. 
The beam energy is 3~GeV, the same as the existing machine. 
Because of the limited cell length, the focusing quadrupole magnets in the arc are made to be combined-function 
magnets to provide bending, too. For example, the center 
focusing magnet has a bending angle of 15.0~mrad and the quadrupole between the 
first and second dipoles (and the fifth and sixth dipoles) has a bending angle of 21.0~mrad. 
The arc focusing quadrupoles have the maximum gradient, at 100 T/m. 
The magnet strengths in this design study are achievable with existing technology as demonstrated 
on the ESRF-EBS project, assuming comparable vacuum pipe radius.  
The other bending magnets are combined-function magnets with defocusing. 
The focusing quadrupoles in the dispersion bump are negative bends, each with a bending angle of 
$-2.8$~mrad. 
The minimum gap between magnets in the lattice is 4~cm. 
The phase advances for the standard cells are 
$\psi_x=1.7917\times2\pi$ and $\psi_y = 0.8426\times2\pi$, respectively. 
The phase advances between the centers of the dispersion bumps (as marked by the center of the SF magnet) in the standard cell are
$\psi_x=0.4929\times2\pi$ and $\psi_y = 0.4889\times2\pi$, respectively.
To minimize the emittance, the horizontal dispersion is set to 7~mm at the center of the ID straight section.
With a total bending angle of $\frac{2\pi}{17}$, the standard cell has an emittance of 
394~pm (short form for pm$\cdot$rad) and a momentum spread of $\sigma_\delta = 1.08\times10^{-3}$. 

\begin{figure}[t]
\includegraphics[width=0.8\textwidth]{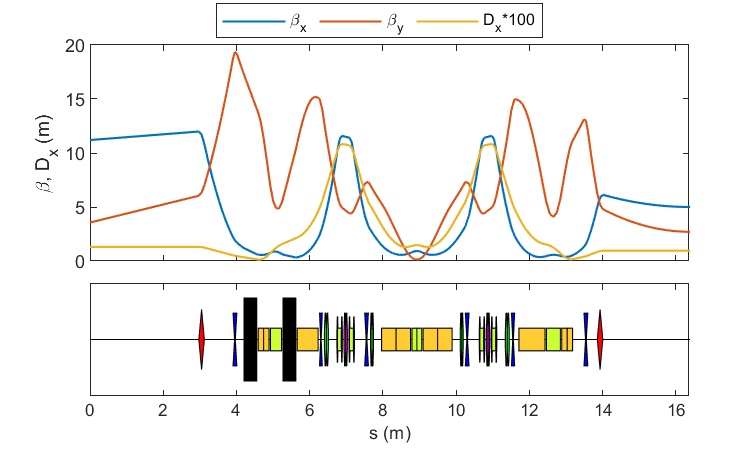}
\caption{The beta and dispersion functions for the east-side matching cell in the SSRLUP lattice. 
The third and fourth kickers are marked with dark rectangles.
\label{figSSRLmatchcellOptics}}
\end{figure}

The matching  cells have a similar H6BA structure (see Figure~\ref{figSSRLmatchcellOptics}). The  long straight section on the east side is reserved 
for injection. The horizontal and vertical beta functions at the injection point are $\beta_x=11.2$~m and 
$\beta_y=3.6$~m, respectively. 
Four injection kickers are placed in the matching cells, which are used to make a closed orbit kicker bump. As the distances between some magnets in the 
east side matching cells are adjusted to accommodate the kickers, the east long straight section is shortened to 6.0~m (all straight sections are measured from quadrupole edges). 
The long straight section on the west side will be used to house the RF cavities and its 
length is 7.2~m. The full ring betatron tunes are $\nu_x=32.250$ and $\nu_y=15.167$. 

Because of the changes in the bending profile throughout the ring, it is impossible to keep all ID 
source points fixed at the original positions. 
However, we are able to keep the source points in one half of the ring fixed. 
The resulting  ring  circumference is 18.5~cm shorter than SPEAR3. 
The emittance for the full  lattice is 364~pm and the momentum spread is 
$\sigma_\delta = 1.07\times10^{-3}$, while the horizontal damping partition is $J_x=1.86$. 
If we include the radiation damping and excitation effects of the existing wigglers  in 
SPEAR3, the emittance becomes 310~pm and the momentum spread 
$\sigma_\delta = 1.04\times10^{-3}$.

\subsection{Nonlinear lattice performance}

The linear lattice fundamentally determines the nonlinear lattice performances. 
While the H6BA lattice cell is generally an ideal cell structure, it is essential to fine tune the lattice parameters for optimal nonlinear dynamics performance. 
The matching of chromatic distortions of the matching cells to the standard cells also plays a critical 
role. 
In the end, the SSRLUP lattice achieves more than a factor of 20 emittance reduction from SPEAR3 while maintaining excellent nonlinear dynamics performance, with 
both DA and Touschek lifetime on par with SPEAR3. 

The natural chromaticities for the standard cell are 
$C_{x0}=-3.79$ and $C_{y0} =-4.11$, respectively, 
while for the full ring the values are $C_{x0}=-69.87$ and $C_{y0} =-74.97$. 
The chromaticities are corrected to $C_{x}=2.7$ and $C_{y} = 0.6$ with 
sextupoles in the dispersion bump area. 
The  sextupole strengths are below $B_2 \equiv \frac{\partial^2B_y}{\partial x^2}= 4200$~T/m$^2$. % (in MAD8 convention). 
No octupole or decapole magnets are used.

\begin{figure}[t]
\includegraphics[width=0.8\textwidth]{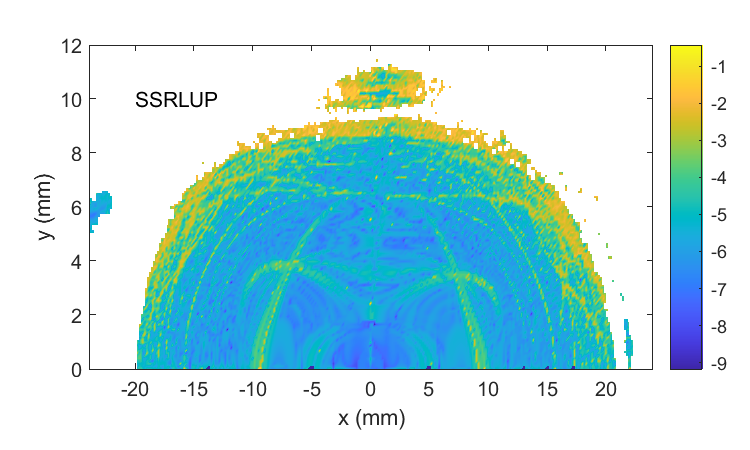}
\includegraphics[width=0.8\textwidth]{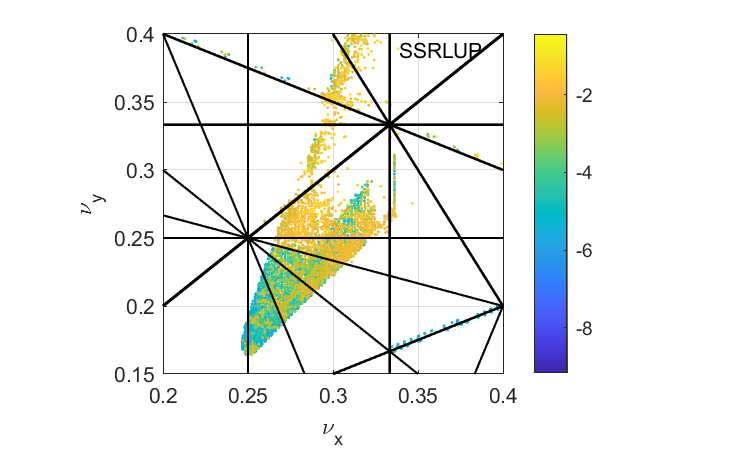}
\caption{Frequency map analysis for the  SSRLUP lattice. 
Top: Tune diffusion  vs. 
launching position at the injection point, where $\beta_x=11.2$~m and 
$\beta_y=3.6$~m; 
bottom: tune diffusion in the tune diagram. 
\label{figSSRLUPxyDetune}}
\end{figure}

\begin{figure}[t]
\includegraphics[width=0.8\textwidth]{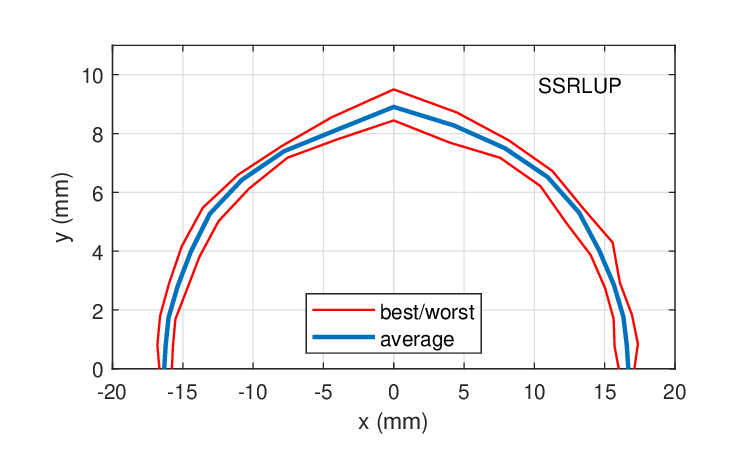}
\caption{The dynamic aperture for SSRLUP, obtained with full 6D particle tracking 
simulation for 5000 turns and with linear optics errors (25 seeds). 
\label{figSSRLUPDA25seeds}}
\end{figure}

Frequency map analysis~\cite{LASKAR1993257naff,FMALaska} was performed for the SSRLUP lattice. 
Particles are tracked for 1024 turns with initial offsets in $x$ and $y$ coordinates 
and the tune diffusion, defined as $\frac12 \text{log}_{10}(\Delta\nu_x^2+\Delta\nu_y^2)$, 
where $\Delta\nu_{x,y}$ are tune changes between the first 512 turns and 
the next 512 turns, are evaluated. 
Figure~\ref{figSSRLUPxyDetune} shows the tune diffusion vs. launching position at the 
injection point (top plot) and in the tune diagram (bottom plot). 
%Particle motion is stable within an elliptical area defined by 
%$x=20$~mm and $y=8.5$~mm. 
There is a very large area of stable motion, which is not plagued by any major 
resonance line. 
The DA is checked with full 6-dimensional (6D) tracking (with radiation 
damping and RF cavities included) for 5000 turns. 
Random linear optics errors are generated in the lattice and are included in tracking simulations, 
with rms beta beating at about $1\%$ (average values over 25 seeds are $1.0\%$ for the horizontal 
plane and $1.3\%$ for the vertical plane). 
Figure~\ref{figSSRLUPDA25seeds} shows the average DA as well as the best and 
worst cases for 25 random seeds. 
The average DA is larger than 16.5~mm in the horizontal plane. 
In this study, we have not yet added nonlinear multipole errors in lattice performance evaluations.


\begin{figure}[t]
\includegraphics[width=0.8\textwidth]{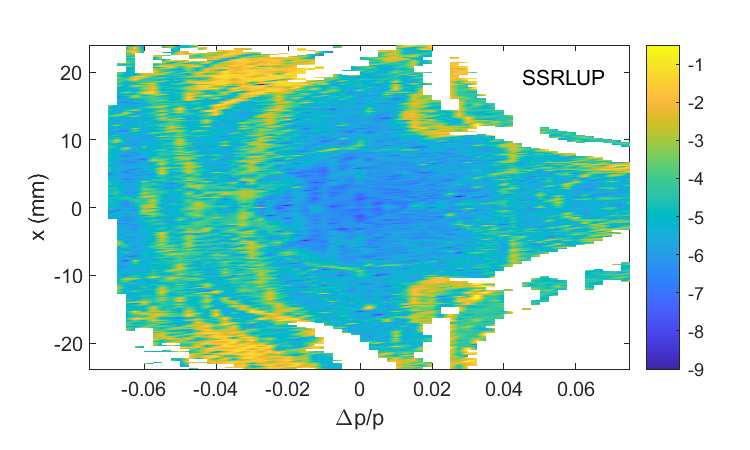}
\includegraphics[width=0.8\textwidth]{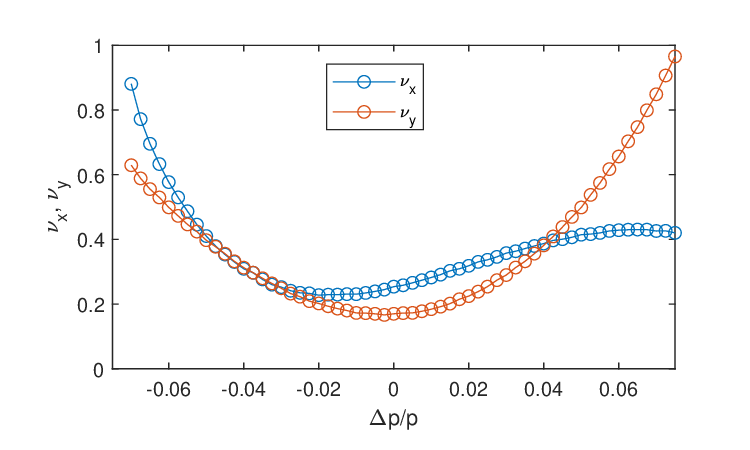}
\caption{Off-momentum frequency map analysis for SSRLUP. 
Top: tune diffusion in ($x$, $\delta=\frac{\Delta p}{p}$) plane; 
bottom: off-momentum betatron tunes. 
\label{figSSRLUPxdppDetune}}
\end{figure}

\begin{figure}[t]
\includegraphics[width=0.8\textwidth]{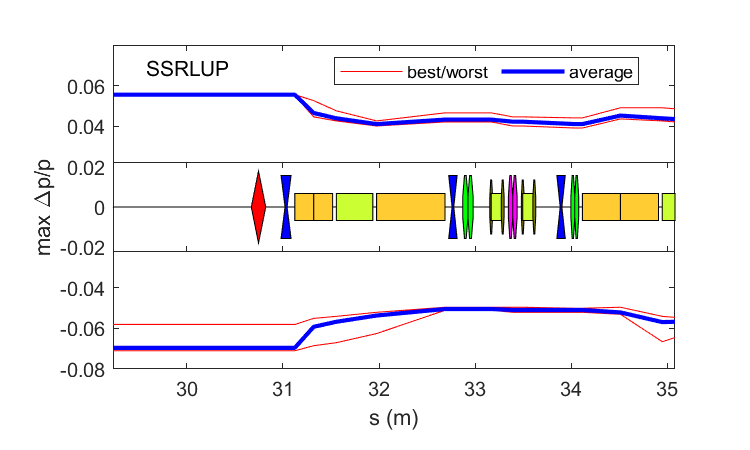}
\caption{The LMA for half of a standard cell for 
SSRLUP. The average, best and worst values of the same 25 seeds as in Figure~\ref{figSSRLUPDA25seeds} are shown. 
\label{figSSRLUPMA25seeds}}
\end{figure}

Stability of off-momentum particles can be studied with frequency map analysis 
in the ($x$, $\delta=\frac{\Delta p}{p}$) plane. 
Figure \ref{figSSRLUPxdppDetune} shows the tune diffusion of particles launched with 
initial $x$ and $\delta$ offsets (top plot) 
and the betatron tunes dependence on momentum deviation for off-momentum particles. 
Stable motion is achieved for a very large range of momentum deviation. 
Full 6D particle tracking is also used to determine the local momentum acceptance (LMA) for a half of 
the standard cell by launching 
particles with initial momentum errors from various locations
(see Figure~\ref{figSSRLUPMA25seeds}). 
 The lattices have the same random optics errors as 
used for DA evaluation. 
In the arc region, the LMA ranges between $-5\%$ and $4\%$. 

The large LMA results in a long Touschek lifetime. 
For the 25 lattices with random errors, the Touschek lifetime for a 500 mA 
beam distributed evenly in 280 bunches is found to be 
between 4.5 and 5.5 hrs, with the average at $4.9$~hrs. 
The calculated Touschek lifetime is close to that of the current SPEAR3 ($5\sim6$ hrs). 
In the calculation, the RF voltage is set to 3~MV, which results in a bunch length of 
$\sigma_z=3.5$~mm. 
The coupling ratio is assumed to be $\epsilon_y/\epsilon_x=3.3\%$, corresponding to a vertical emittance of 10~pm.

\section{SSRL-X: The 587-m lattice}
\subsection{Linear lattice}
A second SSRL upgrade option is to build a new medium energy storage ring 
in a new location on the SLAC site. 
This ring can have a larger  circumference than the SPEAR tunnel, which helps approach  diffraction limited 
performance as well as allows for a larger experimental hall with longer beamlines than SSRLUP. It was determined that a ring with a circumference of 600~m can be fit to the 
available space. 
In the design study process, the beam energy was set to 4~GeV for this new ring by 
considerations of photon energy reach and minimizing the detrimental effects from intra-beam 
scattering (IBS). 
In the following this ring is referred to as SSRL-X. 

SSRL-X consists of a total of 44 H6BA cells. The cells are not identical. 
The center section of the cells, between the sextupoles in the dispersion bumps, are the same. 
However, the end sections are different. 
The phase advances between the centers of the dispersion bumps (as marked by the center of the SF magnet) are
$\psi_x=0.4818\times2\pi$ and $\psi_y = 0.4894\times2\pi$, respectively.
There are 5 different types of end sections, which are  placed with mirror symmetry with the adjacent 
cells, creating 5 types of insertions. 
There are 10 super-bend insertions, each with a 4-mrad bending angle and a maximum bending field of 2~T. 
Another 10 insertions are 1-m long short straight sections, which can host short insertion devices or 
diagnostic equipment.
Interleaved between these are 20 standard straight sections, each is 3.8 m long.
There is one 12-m long straight section reserved for injection. 
On the opposite side of the injection straight, there is an additional standard straight section. 
The two straight sections adjacent to this standard straight section are 3.2 m long. 
There are 21 standard straight sections in total, which are to host 
the primary insertion devices. 
%No negative bend is employed in this lattice. 
The ring has mirror symmetry about the line connecting the injection point to the 
 center of the standard straight section on the opposite side. 
On each side between these two points, the two types of double H6BA cells, one with super bend at the center the other 
with the 1-m short straight, are placed alternately. 
Therefore, on one side of the ring, starting from the injection straight, we have a super bend, 
a standard straight, a 1-m straight, a standard straight, and so on, until it reaches the 3.2-m 
straight and then the standard straight in the opposite position. 

\begin{figure}[t]
\includegraphics[width=0.8\textwidth]{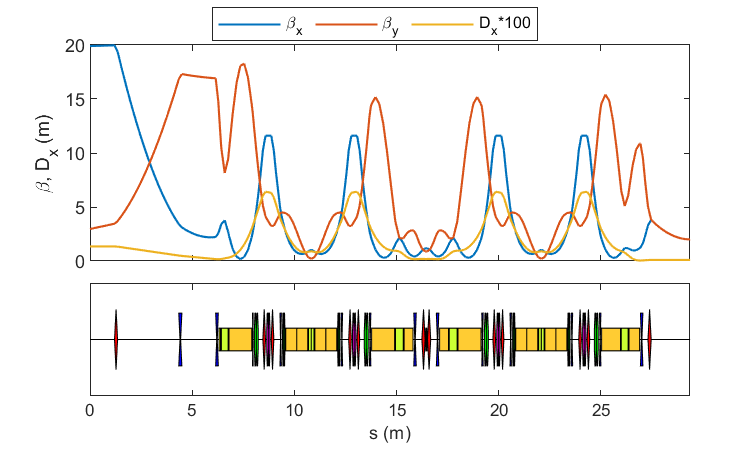}
\includegraphics[width=0.8\textwidth]{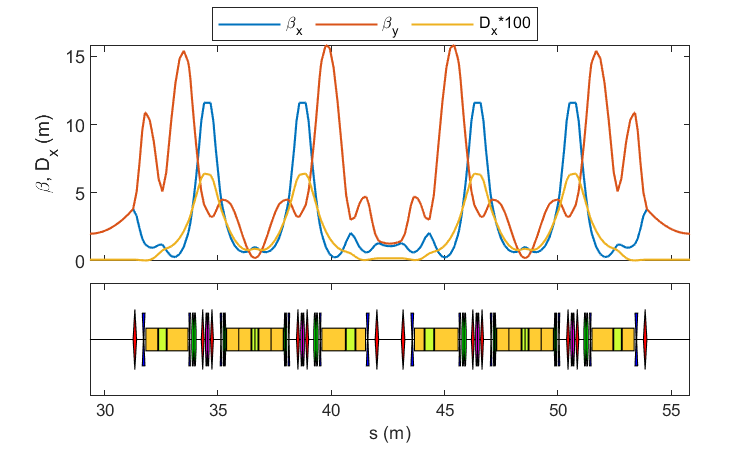}
\caption{The linear optics functions for SSRLX lattice cells. 
Top: two H6BA cells consisting of half of the injection straight, one superbend (at $s=16.45$~m), and half a standard straight;
bottom: two H6BA cells with a 1-m short straight in the middle, and two half standard straights. 
\label{figSSRLXoptics}}
\end{figure}

Figure~\ref{figSSRLXoptics} shows the linear optics functions of the double H6BA cells. 
The top plot also shows a half of the injection long straight section. 
The beta functions at the injection point are $\beta_x=20$~m and $\beta_y=3.0$~m, respectively. 
The beta functions at the center of the standard ID straight sections are 
$\beta_x=\beta_y=2.0$~m, which are optimal values for matching to the photon beams for the ID lengths. 

The full ring lattice has a circumference of 586.7~m. The natural emittance is $85.7$~pm and the 
rms momentum spread is $\sigma_\delta = 0.67\times10^{-3}$. The horizontal damping partition is 
$J_x=1.41$. 
The betatron tunes are $\nu_x=78.2$ and $\nu_y = 37.2$, respectively. 
If damping wigglers of a total length of 20~m, with a peak field of 1~T and a period of 124~mm, are installed, the emittance decreases to 69.2~pm, while the momentum spread increases to $\sigma_\delta = 0.73\times10^{-3}$.
%If damping wigglers of a total length of 35~m, with a peak field of 1~T and a period of 124~mm, are installed, the emittance decreases to 60.5~pm, while the momentum spread increases to $\sigma_\delta = 0.76\times10^{-3}$. 
%If the  damping wiggler length is 20~m in total, the emittance becomes 69.2~pm and the momentum spread $\sigma_\delta = 0.73\times10^{-3}$. 

\subsection{Nonlinear lattice performance}
The natural chromaticities for the SSRL-X lattice are $C_{x0}=-155.0$ and $C_{y0}= -139.5$ 
respectively. These are corrected to low positive values with sextupoles. The 
sextupole strengths ($B_2$) are below $4400$~T/m$^2$.

Because of different types of H6BA cells, the ring has only a super-periodicity of one. 
However, the H6BA cells are arranged in pairs with mirror symmetry and each pair, 
even consisting of different types, has identical betatron phase advances. 
In addition, sextupole magnets in the different types of cells have the same beta function values  and 
the two main types of double cells are placed alternately. 
Therefore, there is some cancellation of nonlinear resonances from the contributions of the cells.  
%Effectively, there is a super-periodicity of 22.

\begin{figure}[t]
\includegraphics[width=0.8\textwidth]{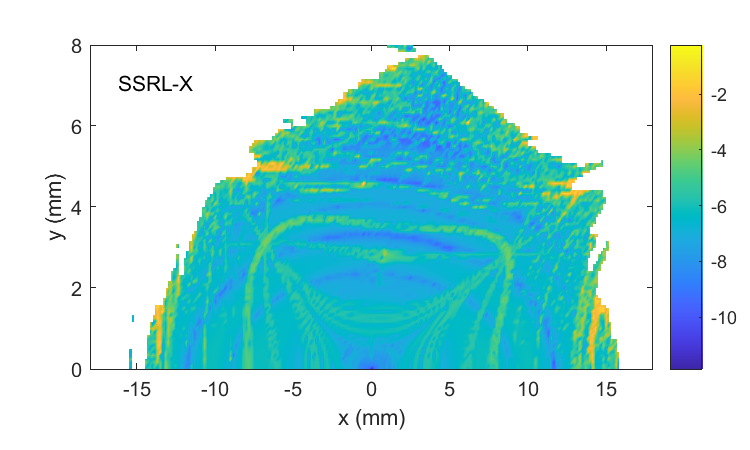}
\includegraphics[width=0.8\textwidth]{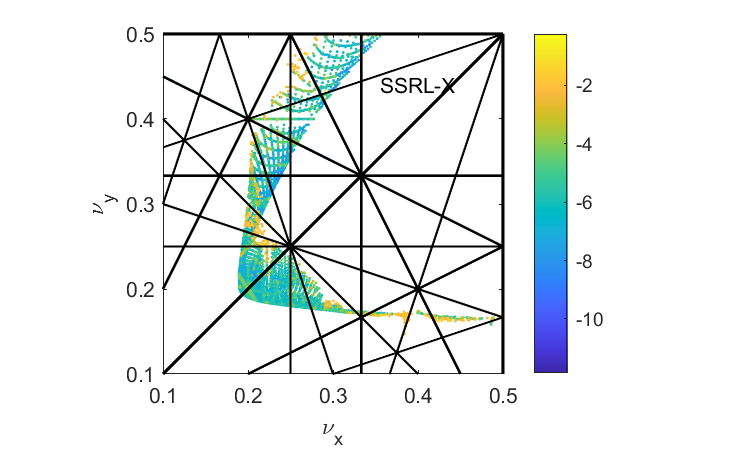}
\caption{Frequency map analysis for the  SSRL-X lattice. 
Top: tune diffusion  vs. 
launching position at the injection point, where $\beta_x=19.9$~m and 
$\beta_y=3.0$~m; 
bottom: tune diffusion in the tune diagram. 
\label{figSSRLXxyDetune}}
\end{figure}

\begin{figure}[t]
\includegraphics[width=0.8\textwidth]{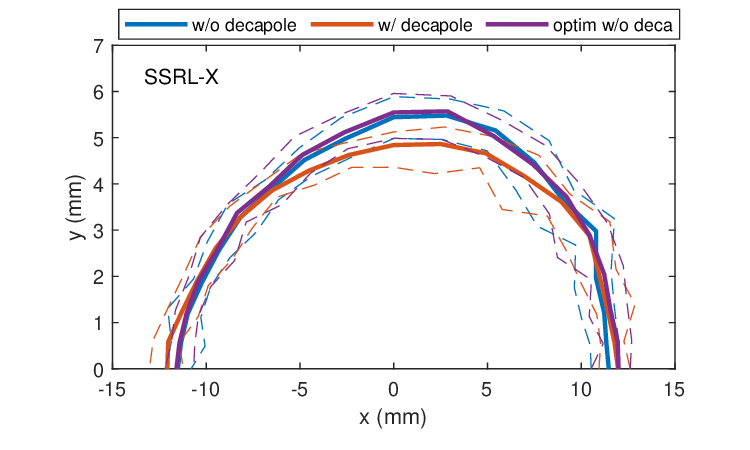}
\caption{ DA evaluated with random linear errors (25 seeds) for SSRL-X, with  
decapoles off (``w/o decapole'') or on (``w/ decapole''). Also shown is the DA
for the selected solution optimized with decapoles off (``optim w/o deca''). 
\label{figSSRLXxyDA}}
\end{figure}

Figure~\ref{figSSRLXxyDetune} shows the frequency map of the SSRL-X lattice, evaluated by launching particles 
from the middle of the injection straight. 
Figure~\ref{figSSRLXxyDA} shows the DA evaluated with random linear errors and by 
tracking 5000 turns. 
The average rms beta beating for the 25 seeds are $1.1\%$ for both planes. Skew quadrupole 
errors are introduced at sextupole locations, resulting in coupling ratio ranging from 
4\% to 100\%, averaging  39\% among the seeds. 
The DA is up to 12~mm in the horizontal plane,  roughly 
corresponding to the condition when the horizontal betatron tune reaching $\nu_x=0.3$ with amplitude-dependent detuning. 
The vertical DA is 5~mm, corresponding to shifted vertical tune of $\nu_y=0.29$. 
Small amplitude-dependent detuning, achieved through fine tuning of the linear knobs in the H6BA 
cells~\cite{Pantaleo2023PRAB}, is critical for the large DA. 
Neither DA nor beam lifetime (to be discussed below) is found to be correlated with the 
coupling ratio in the random seeds. 

\begin{figure}[t]
\includegraphics[width=0.8\textwidth]{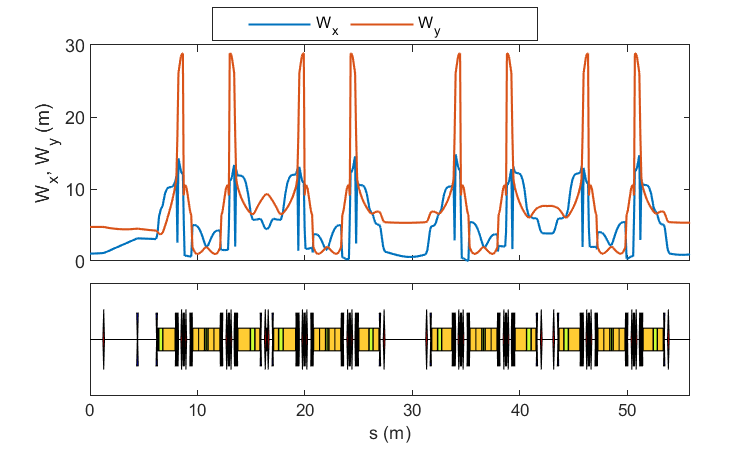}
\caption{The horizontal and vertical Montague functions describing achromatic aberration 
for the first 4 H6BA cells of SSRL-X.
\label{figSSRLXWxWy}}
\end{figure}

\begin{figure}[t]
\includegraphics[width=0.8\textwidth]{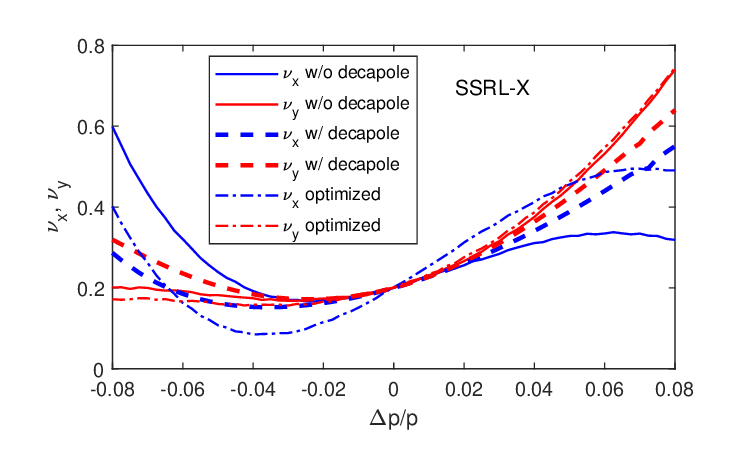}
\caption{Betatron tunes vs momentum deviation for SSRL-X, with or without decapoles. 
The selected optimal solution from the next subsection is also included. 
\label{figSSRLXNuxNuyvsDpp}}
\end{figure}

For the motion of off-energy particles, the transparency conditions require the Montague functions~\cite{Montague} to be 
matched between different sections. 
Figure~\ref{figSSRLXWxWy} shows these functions for the first 4 H6BA cells, starting from the injection point. 
There is no global distortion as indicated by the pseudo-periodicity across the different types of cells. 
Figure~\ref{figSSRLXNuxNuyvsDpp} shows the betatron tune dependence on the momentum deviation within the 
$\pm8\%$ range. The horizontal and vertical betatron tunes cross the half integer resonance on the negative and 
positive ends, respectively, due to third order chromaticities. 
A family of 88 decapoles located in the focusing sextupole (SF) magnets, each with an integrated strength of 
$K_4L\equiv \frac{L}{B\rho}\frac{\partial^4B_y}{\partial x^4}=
1.6\times10^5$~m$^{-4}$, is used to correct the third order chromaticities. 
The  tunes vs. momentum deviation curves for the case with decapoles are also included in 
Figure~\ref{figSSRLXNuxNuyvsDpp}. 
Without the decapoles, the third order chromaticities are $C_{x3}=-664$ and $C_{y3}=-169$, respectively, 
which become $C_{x3}=-142$ and $C_{y3}=-52$ when decapoles are included. 
No octupole magnets are used in this lattice. 
%without Decapoles
% >> [px,s] = polyfit(a(1).x, a(1).y,3)
% px = -664.2174   38.5908    2.5233    0.1946
% >> [py,s] = polyfit(a(2).x, a(2).y,3)
% py =  169.1694   42.7887    2.2614    0.2043

%with decapoles
% >> [px,s] = polyfit(a(1).x, a(1).y,3)
% px = -142.1043   33.1336    2.5916    0.1965
% >> [py,s] = polyfit(a(2).x, a(2).y,3)
% py =  -52.4842   44.1689    2.3210    0.2019


\begin{figure}[t]
\includegraphics[width=0.8\textwidth]{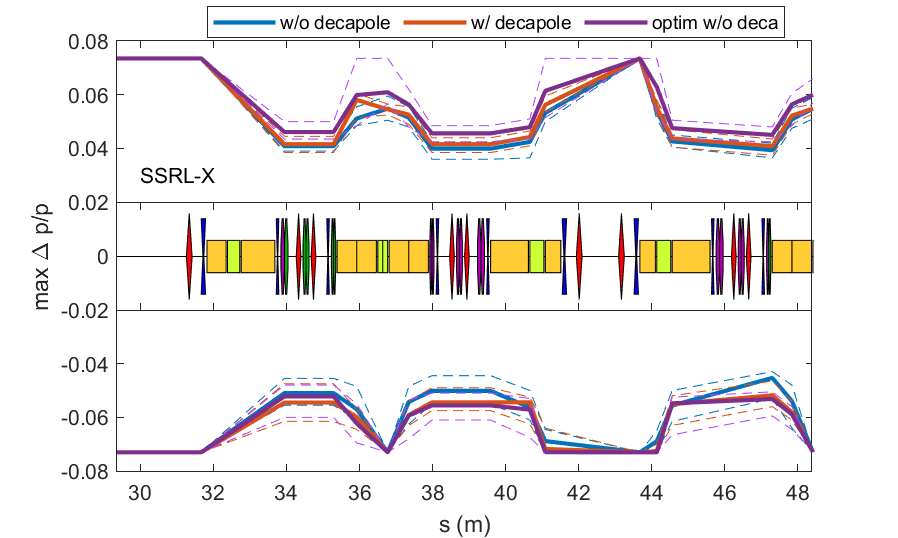}
\caption{Momentum aperture for half of an SSRL-X double H6BA cell (with 1-m middle straight), with (``w/ decapole'') or without (``w/o decapole'') decapoles, 
evaluated with linear errors in the lattice. 
Solid curves are average values, while dashed curves represent best and worst values of 25 seeds. 
Also shown are the MA for the selected solution from optimization without decapoles (``optim w/o deca''). 
\label{figSSRLXcmpMADeca}}
\end{figure}

With the correction of third order chromaticities by the decapoles, the MA  
is improved, as shown in Figure~\ref{figSSRLXcmpMADeca}. 
The corresponding calculated Touschek lifetime are compared in Figure~\ref{figSSRLXcmpTLT},   
for a bunch current of $I_b=0.57$~mA (400 mA in 700 bunches). 
A  coupling ratio of $\frac{\epsilon_y}{\epsilon_x}=20\%$ is assumed. 
The bunch length is set to $\sigma_z=9$~mm in the 
calculation, which corresponds to an rf voltage of 3~MV and with bunch lengthening by a factor of 4 by a harmonic cavity. 
The Touschek lifetime is improved from 29.8 hr to 34.4 hr (mean value for 25 seeds) with the decapoles. 
The introduction of decapoles modifies the tune footprint for on-energy particles. 
The DA remains about the same on the horizontal plane, but shrinks from 5.6~mm to 4.8~mm in the 
vertical plane (see Figure~\ref{figSSRLXxyDA}). 

\subsection{Nonlinear dynamics optimization}
The decapoles are currently modeled as thin elements sandwiched at the center of the SF  
sextupole  magnets. 
Although the decapoles  help the off-momentum beam dynamics, 
it is preferable not to include them since adding the decapole component on the SF magnets complicates the magnet design. In order to achieve similar nonlinear dynamics performance without the decapoles, we 
performed sextupole optimization, using the multi-objective evolutionary algorithm particle 
swarm optimization (MOPSO) as used in Ref.~\cite{HUANG201448}. 
The DA area and the Touschek lifetime are the two objective functions. 
The DA area is evaluated with twice the weight for the part with $x<0$ as this is the region to accept the injected beam. The LMA at 18 monitor points over half a H6BA cell are evaluated 
and used for Touschek lifetime calculation.

Six sextupole knobs are used as optimization parameters, which are formed as follows. 
The ring lattice  can be considered as 22 double H6BA cells. Although not all these double cells are the same, 
they have the same sextupole magnets. 
We require the sextupoles  at  the same locations in the double cells to have the equal strengths. 
There are 12 sextupole magnets in each double cell. 
By requiring mirror symmetry, they can be grouped into six pairs; each pair makes one knob. 

\begin{figure}[t]
\includegraphics[width=0.8\textwidth]{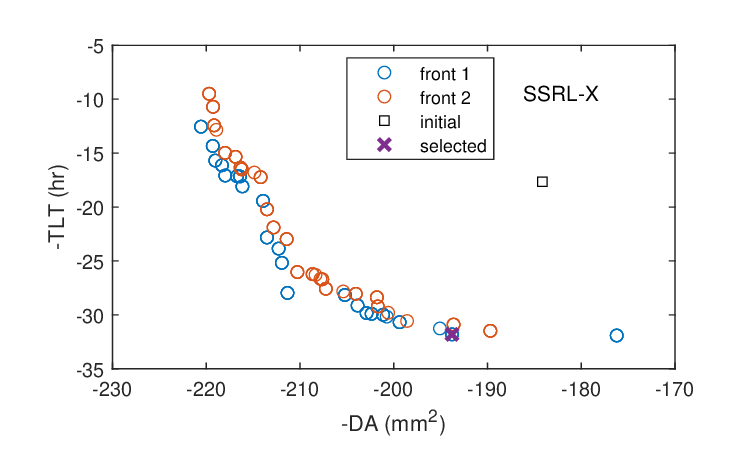}
\caption{The objective functions of the solutions in the leading fronts obtained by non-dominated 
sorting in the SSRL-X sextupole optimization.  
Note the Touschek lifetime (``TLT'') is calculated with a different setting in  bunch current ($I_b=0.41$ mA), coupling ratio ( $\frac{\epsilon_y}{\epsilon_x}=100\%$), bunch length ($\sigma_z=2.2$~mm) than with the 
full evaluation shown in Figure~\ref{figSSRLXcmpTLT}. 
\label{figSSRLXsextoptimLF}}
\end{figure}

After 6000 solutions are evaluated, the two objectives are improved in  the final non-dominated solutions, as seen in Figure~\ref{figSSRLXsextoptimLF}. 
In nonlinear lattice optimization, the lifetime calculation was based on a bunch with full coupling and without 
bunch lengthening. 
At a later stage, we changed the coupling ratio to 20\% for better photon beam brightness, for which bunch lengthening is desired to alleviate IBS effect (see section~\ref{secBrightness}) . 
A solution with a significant improvement in Touschek lifetime is selected for full evaluation with linear optics errors. 
The chromaticities of the selected solution are $C_x =5.3$ and $C_y =2.6$, up from $C_x=C_y=2.5$ for the 
original solution. 

\begin{figure}[t]
\includegraphics[width=0.8\textwidth]{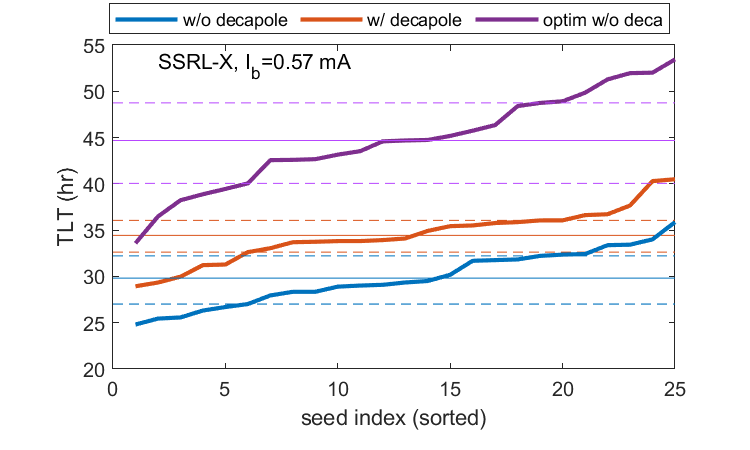}
\caption{Touschek lifetime for SSRL-X beam evaluated for bunch current of 0.57 mA, for the original sextupole setting with (``w/ decapoles'') or without (``w/o decapoles'')
decapoles, or for the optimized solution without decapoles (``optimized w/o deca''). Results for the same 25 random linear error seeds are shown. Solid horizontal lines show the 
average values, while the dashed lines indicate the 25\% and 75\% percentiles. 
Coupling ratio of $\frac{\epsilon_y}{\epsilon_x}=20\%$ and 
bunch length of $\sigma_z=9$~mm (w/ bunch lengthening and IBS) are assumed.
\label{figSSRLXcmpTLT}}
\end{figure}

The DA of the selected solution is similar to that of the original solution without decapoles, 
as shown in Figure~\ref{figSSRLXxyDA}. 
The LMA of the selected solution is shown in Figure~\ref{figSSRLXcmpMADeca} in comparison with the original lattice, with or without 
decapoles. 
A comparison of the Touschek lifetime for the three cases are shown in Figure~\ref{figSSRLXcmpTLT}. 
As the result of the improvement in LMA, the Touschek lifetime is  increased to 44.7~hr, which is 
even better than the case with decapoles. 

\section{SDLS: 2190-m ring}
Building a new storage ring in the PEP tunnel as an upgrade for SSRL has been previously considered in multiple efforts~\cite{hettel2008ideas,HettelPAC09,WANG201130,cai2012ultimate}. 
In this study, a lattice based on the H6BA cell is designed for the same ring geometry for a 5-GeV beam, which we 
refer to as Stanford Diffraction Limited Synchrotron (SDLS). 

\subsection{Linear optics}
\begin{figure}[t]
\includegraphics[width=0.8\textwidth]{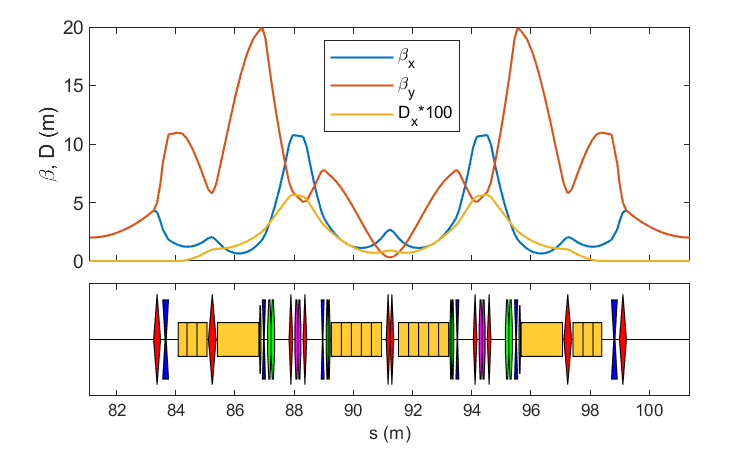}
\includegraphics[width=0.8\textwidth]{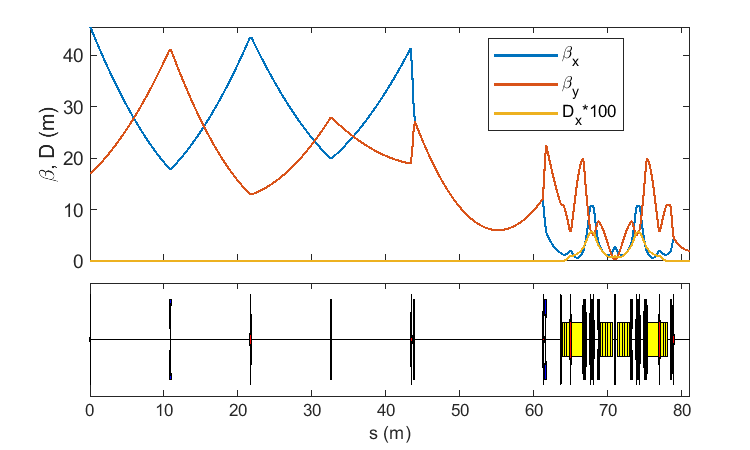}
\caption{The linear optics functions for the SDLS lattice. 
Top: One H6BA cell, with phase advances 
$\psi_x=1.7667\times2\pi$ and $\psi_y = 0.9056\times2\pi$, respectively.  
bottom: Half of the long straight section for injection and two H6BA cells. 
\label{figSDLSoptics}}
\end{figure}

The PEP tunnel has a 6-fold geometry, with 6 arcs and 6 long straight sections. As the length of one long straight section is 121.5~m, out of the circumference of 2190~m, the total arc length available to host H6BA cells is 1461~m. 
Each of the six arcs consists of 12 identical H6BA cells, with cell length equal to 20.3~m.
The standard straight section has a length of 4.3~m, with the horizontal and vertical beta function 
both at $2.0$~m. 
The betatron phase advance for one H6BA cell is $\psi_x=1.7667\times2\pi$ and $\psi_y = 0.9056\times2\pi$, respectively. 
The phase advances between the centers of the dispersion bumps (as marked by the center of the SF magnet) are
$\psi_x=0.4732\times2\pi$ and $\psi_y = 0.4634\times2\pi$, respectively.
The bare lattice emittance is 28.3~pm, while the  momentum spread is 
$\sigma_\delta = 0.63\times 10^{-3}$. The horizontal damping partition is $J_x=1.67$. 
The momentum compaction factor is $\alpha_c=2.6\times10^{-5}$. 
The long straight section optics is configured for off-axis injection, with  the  beta 
functions increased to $\beta_x=45.5$~m and $\beta_y=17.0$~m, respectively. All six long straight sections have the same optics, although the ones 
other than the actual injection section can be configured differently. 
Figure~\ref{figSDLSoptics} shows the linear optics functions of the H6BA cell (top plot)  
and half of the long straight section with one H6BA cell (bottom plot). 

Since the PEP tunnel is underground, excavation is needed to create space for photon beamlines.
In previous work it has been determined that only two of the six arcs are suitable for excavation and for building beamlines~\cite{HettelPAC09}. Therefore, there are spaces to build 22 beamlines with IDs in standard straight sections. Four additional beamlines can be built with IDs in the adjacent long straight sections. 
More IDs can be potentially hosted in the long straight section with canting.  

Damping wigglers can be installed in the standard straights sections in the arcs not planned for beamlines. Or they can 
be placed in the long straight sections. 
If damping wigglers of a total length of 70-m are employed, with the  peak field of 1~T and wiggler period of 124~mm, 
the beam emittance decreases to 15.0~pm,  
the momentum spread becomes $\sigma_\delta = 0.88\times 10^{-3}$, and 
the horizontal damping partition is $J_x=1.257$. 
The one-turn radiation energy loss becomes 1.79~MeV, up from 0.68~MeV for the bare lattice. 

%Therefore, the standard straight sections in the other arcs can be used to host damping wigglers. 
%One straight section can host two 2-meter long damping wigglers, each with 16 periods, with the wiggler period of 124~mm and peak field of 0.6~T (corresponding to 5~cm half poles with bending radius of 35.7~m when the poles are modeled as short dipoles). 
%integraged L/rho^2=(1.4e-3/.05)^2*0.1=7.8400e-05 per period, peak field =sqrt(7.8400e-05*2/0.124)*50/3=0.593 T
%With 88 damping wigglers, the beam emittance decreases to 15.5~pm,  
%the momentum spread becomes $\sigma_\delta = 0.69\times 10^{-3}$, and 
%the horizontal damping partition is $J_x=1.28$. 
%The one-turn radiation energy loss becomes 1.65~MeV, up from 0.68~MeV for the bare lattice. 

\subsection{Nonlinear dynamics \label{secSDLS:nl}}
The natural chromaticities for the bare lattice are  
$C_{x0}=-199.3$ and $C_{y0}= -213.8$. 
These are corrected to $C_{x}= 6.3$ and $C_{y}= 9.1$, respectively, 
with three families of sextupole in the dispersion bump area. 
The sextupole strengths ($B_2$) are below
3700~T/m$^{2}$. % (MAD8 convention). 

One family of octupoles, located in the dispersion bump region just downstream of the second bending magnet and its 
mirror symmetry point in the H6BA cell,
where the vertical beta function is high,  is employed to control the amplitude dependent detuning. 
The octupole is modeled as a 5-cm long element in the lattice. With its strength set at
$B_3\equiv\frac{\partial^3 B_y}{\partial x^3}=-16200$~T/m$^{3}$, it substantially decreases the cross-term amplitude dependent detuning coefficient,  $\frac{d\nu_x}{dJ_y}=\frac{d\nu_y}{dJ_x}$, while also changing the sign of 
the $\frac{d\nu_y}{dJ_y}$ coefficient (see Table~\ref{tab:SDLSADTS}). 
This is beneficial for avoiding crossing the integer below by the vertical tune at small horizontal oscillation amplitude. 
The octupoles cause a small increase of the second order chromaticity for the vertical plane. 
\begin{table}[htbp] 
\begin{center} 
\caption{SDLS amplitude dependent detuning coefficients ($\times10^3$) and 
second order chromaticities, with or 
without the octupole family}
\label{tab:SDLSADTS} 
\begin{tabular}{c|c|c|c|c|c} 
\hline
Octupole status & $\frac{d\nu_x}{dJ_x}$ & $\frac{d\nu_x}{dJ_y}=\frac{d\nu_y}{dJ_x}$ & $\frac{d\nu_y}{dJ_y}$ & $C_{x,2}$ & $C_{y,2}$\\
\hline 
Off & $15.9$ & $-146.0$ &$ 40.0$ & 59.4 & 170.6 \\ 
On &  $12.9$ &$-68.2$ &$ -63.3$ & 55.1 & 223.2\\ 
%Optim & $16.0$ & $$
\hline 
\end{tabular} 
\end{center} 
\end{table} 
%octupole off: 
% dnudxy =   1.0e+05 *
%     0.1590   -1.4596
%     --        3.9947
%octupole on:  
% dnu/dJ = 1.0e+04 *
%    1.2908   -6.9472
%   -6.7132   -6.3252

\begin{figure}[t]
\includegraphics[width=0.8\textwidth]{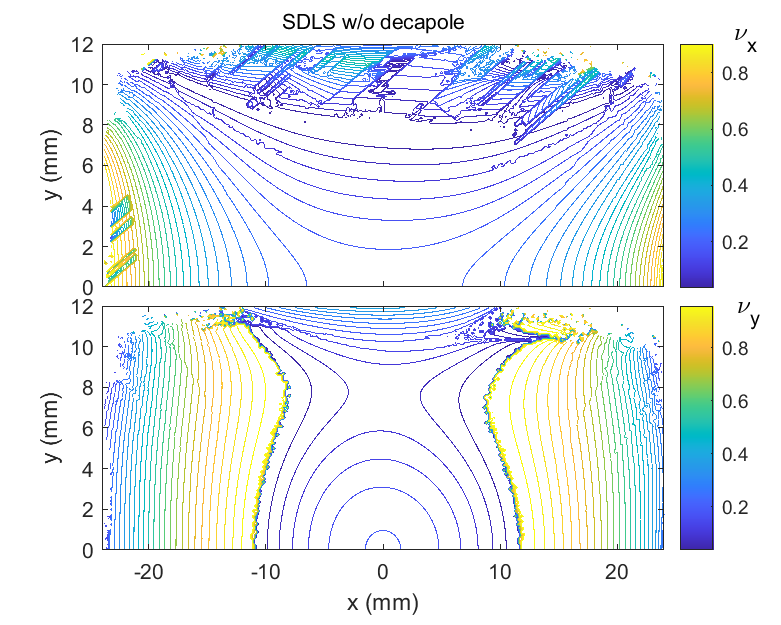}
\includegraphics[width=0.8\textwidth]{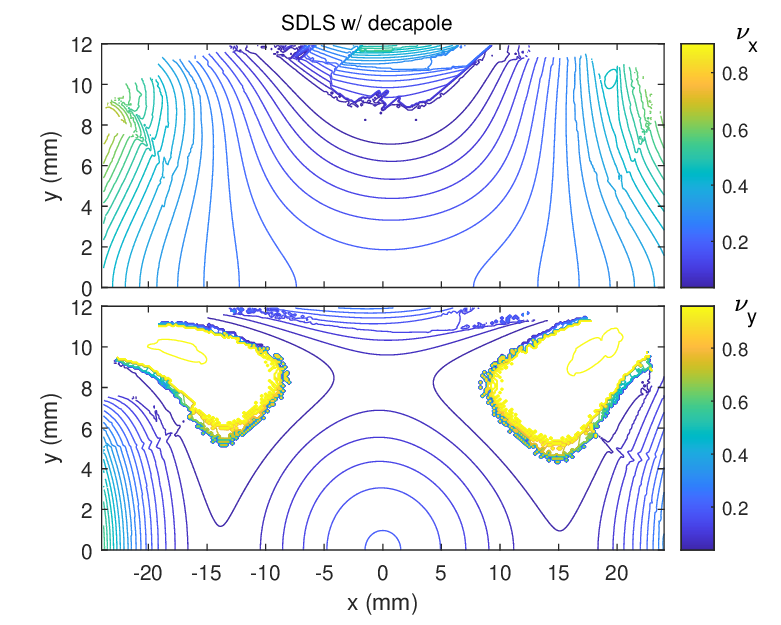}
\caption{The betatron tunes ($\nu_x$ and $\nu_y$) vs. launching position ($x$, $y$) in contour plots 
for SDLS with (bottom plots) or without (top plots) decapoles. 
\label{figSDLScmpTuneXYSDLS}}
\end{figure}

A decapole family is placed near the center of the dispersion bump where the horizontal beta is high 
to help shape the tune footprint of the injected particles and control third order chromaticities. 
It is currently modeled as a thin-element multipole. 
The integrated strength is set to $K_4L=2\times10^5$~m$^{-4}$. 
Figure~\ref{figSDLScmpTuneXYSDLS} shows the contour plots of betatron tunes vs. 
the launching position in the $x$-$y$ plane, with or without the decapoles. 
Without decapoles, the vertical tune crosses the integer from below at about $x=\pm 11$~mm. 
With the decapoles, the vertical tune is turned around before reaching the lower integer as the 
horizontal oscillation amplitude is increased. It crosses the integer only when there is also a 
large vertical offset from the middle plane. 

\begin{figure}[t]
\includegraphics[width=0.8\textwidth]{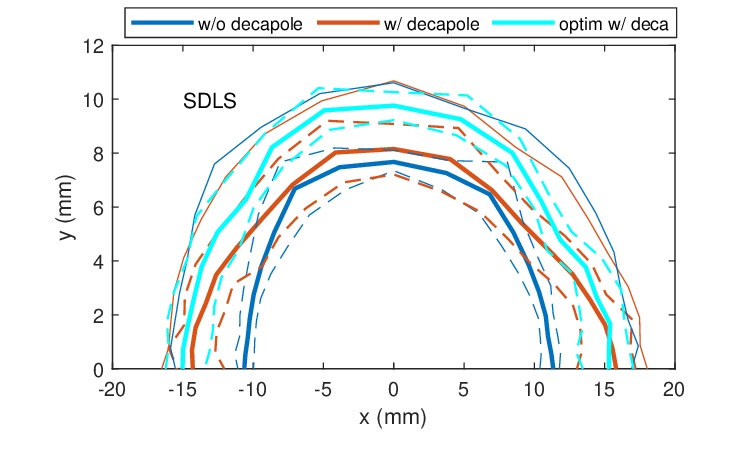}
\caption{The DA for SDLS with (orange) or without (blue) decapoles are compared. The thin solid lines are 
without lattice errors. The thick solid lines are the average of 25 random seeds, while the 
dashed lines are by connecting the best or worst points of the seeds. 
Also shown is the DA for an optimized solution (see section~\ref{secSDLSoptim}). 
The DA is evaluated at the injection point where $\beta_x=45.5$~m and $\beta_y=17.0$~m. 
\label{figSDLScmpDAnoDecavswDeca}}
\end{figure}

Without linear optics errors in the lattice, the DA (obtained by 6D tracking with radiation damping) for the cases with or without the decapole family are similar (see Figure~\ref{figSDLScmpDAnoDecavswDeca}). 
Tracking simulations are also done with random linear optics and coupling errors added to the lattice. 
For the 25 error seeds, the rms beta beating ranges from $0.7\%$ to $2.2\%$ (with a mean value of $1.4\%$) in the horizontal plane and  from $1.2\%$ to $4.3\%$ (with a mean value of $2.5\%$) in the vertical plane. 
By adding skew quadrupole 
errors to sextupole magnets, the coupling ratio for the seeds ranges from 
6\% to 100\%, with an average value of  56\%. 
With the linear errors, the DA for the case without decapole 
collapses to about $\pm 11$~mm in the horizontal plane. 
However, with decapoles, the DA (for the same error seeds) remains at about 15~mm in the horizontal direction. Clearly, it is advantageous to 
shape the tune footprint with decapoles.

% >> n = 33+([-14:14]);
% >> [px0] = polyfit(a(1).x(n), a(1).y(n),3)   %noDeca
% px0 =   1.0e+03 *
%    -1.9898    0.0582    0.0063    0.0002

% >> [px1] = polyfit(a(3).x(n), a(3).y(n),3) %with deca
% px1 =   1.0e+03 *
%    -1.4230    0.0584    0.0063    0.0002

% >> [px2] = polyfit(a(5).x(n), a(5).y(n),3) %optim, id 310
% px2 =   1.0e+03 *
%    -2.0481    0.0627    0.0062    0.0002

% >> [py0] = polyfit(a(2).x(n), a(2).y(n),3) %noDeca
% py0 =  300.8456  227.5834    9.0884    0.2008
% >> [py1] = polyfit(a(4).x(n), a(4).y(n),3)   %with deca
% py1 =  -23.7660  226.0481    9.1108    0.2006
% >> [py2] = polyfit(a(6).x(n), a(6).y(n),3)  %optim, id 310
% py2 =  -86.6151  207.5379    9.4426    0.2001
  
\begin{figure}[t]
\includegraphics[width=0.8\textwidth]{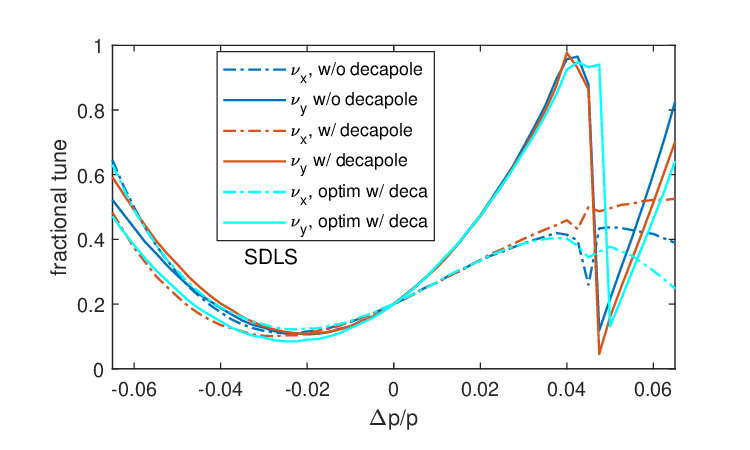}
\caption{Betatron tunes vs. momentum deviation for SDLS, with or without the decapole family. 
Also shown is that of a selected solution from the optimization (see section~\ref{secSDLSoptim}). The apparent sudden drop of the vertical tune beyond $+4\%$ is due to aliasing as the tune crosses the integer. 
\label{figSDLScmpTunesVsDpp}}
\end{figure}

\begin{figure}[t]
\includegraphics[width=0.8\textwidth]{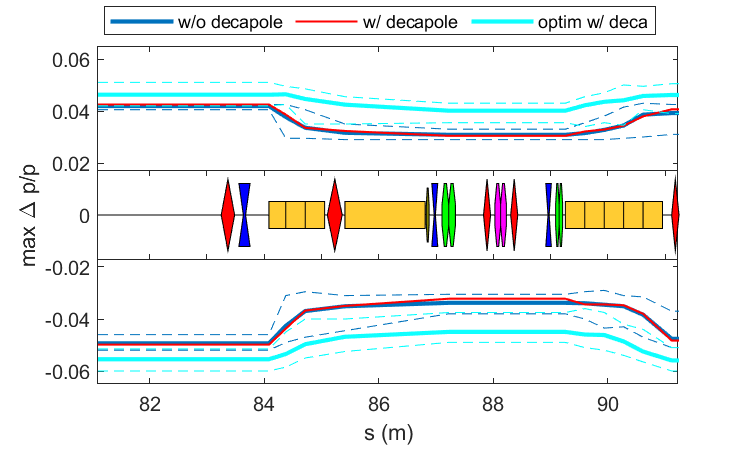}
\caption{The MA in half of an H6BA cell for SDLS with (red) or without (blue) decapoles are compared. 
Also shown are MA for a selected solution from optimization. 
\label{figSDLScmpMAnoDecavswDeca}}
\end{figure}

\begin{figure}[t]
\includegraphics[width=0.8\textwidth]{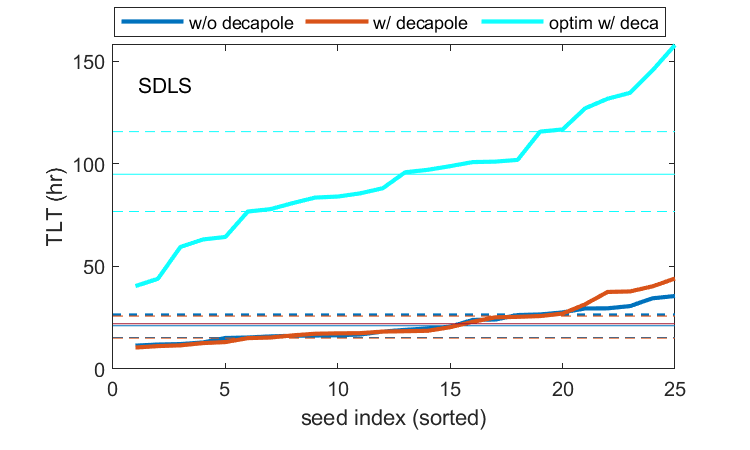}
\caption{The calculated Touschek lifetime for the SDLS lattice with 25 random error seeds
with (red) or without (blue) decapoles are compared. 
Also shown are MA for a selected solution from optimization. 
\label{figSDLScmpTLTnoDecavswDeca}}
\end{figure}

The decapoles  increase the third order horizontal chromaticity (from $C_{x3}=-2000$ to 
 $-1420$)
 and decreases the third order vertical chromaticity (from 
 $C_{y3}=300$ to 
 $-20$)  (see Figure~\ref{figSDLScmpTunesVsDpp}). 
Because of the large second order vertical chromaticity, the vertical tune crosses the integer 
above at the momentum deviation of $4\%$, and the decapoles have little impact on it. 
%The horizontal tune crosses the integer from above at about $-7.3\%$ if the decapole family is turned off and the crossing is delayed by the decapoles. 
If there are no lattice errors in the ring, the case with decapole has a much larger LMA than the case without decapoles. The Touschek lifetime would reach 306 hours for a 0.1 mA bunch with full coupling,  
$\epsilon_x=\epsilon_y=9.6$~pm (w/ IBS), and bunch length of $\sigma_z = 6.4$~mm (w/ bunch lengthening), while the Touschek lifetime is 177 hours for the same conditions except with decapoles turned off. 
However, with the same 25 random error seeds of lattice errors, tracking simulation finds the LMA to be 
about the same with or without the decapoles (see Figure~\ref{figSDLScmpMAnoDecavswDeca}), 
and both are substantially lower than the ideal lattices. 
Consequently, the calculated mean Touschek lifetime for the same conditions is 21.0~hr without decapoles 
and 22.0~hr with decapoles (see Figure~\ref{figSDLScmpTLTnoDecavswDeca}). 


\subsection{Nonlinear dynamics optimization\label{secSDLSoptim}}
The DA and Touschek lifetime for the SDLS as shown in the previous subsection would be adequate for storage ring operation. 
However, in view of the substantial decrease of the Touschek lifetime when lattice errors are included, 
numerical optimizations were carried out to improve the nonlinear dynamics performance under more realistic 
conditions. A set of linear  errors is included in the lattice. 
Similar to the SSRL-X case, the weighted DA area and the Touschek lifetime are the two objective functions.

Sixteen sextupole knobs and the strengths of the octupole and decapole families are used as optimization 
parameters, totaling 18 knobs. 
By requiring the sextupole strengths in an H6BA cell to have mirror symmetry, each cell contributes 3 
sextupole parameters. We further require the sextupole parameters in the 12 cells in one arc to have mirror 
symmetry about the center of the arc. With a 6-fold periodicity, the whole lattice has 18 sextupole parameters. 
Using singular value decomposition, we obtain the 
16 basis vectors of the null space of  the chromaticity response matrix of these 18 parameters and 
use them as the sextupole knobs~\cite{HuangDA2015}. 

\begin{figure}[t]
\includegraphics[width=0.8\textwidth]{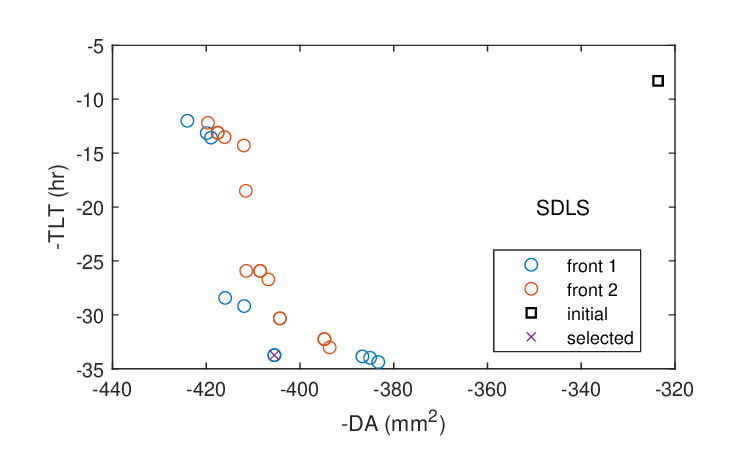}
\caption{The objective functions of the two leading fronts of all evaluated solutions during 
nonlinear dynamics optimization are compared to that of the initial lattice for SDLS. 
Note Touschek lifetime evaluation here is without IBS and without bunch lengthening by 
harmonic cavity. The bunch current is assumed to be $I_b=0.1$~mA. 
\label{figSDLSFront12vsInit}}
\end{figure}

The MOPSO optimization algorithm ran 76 generations, with a population size of 60. 
The objective functions of the best solutions, obtained by non-dominated sorting, 
are compared to that of the initial lattice in 
Figure~\ref{figSDLSFront12vsInit}. One of the leading solutions, which has  
a substantial increase in both DA and Touschek lifetime, is selected for further evaluation 
with multiple lattice error seeds. 
The DA, MA, and Touschek lifetime of this lattice solution are compared to the original lattice, 
with or without decapoles, in Figures~\ref{figSDLScmpDAnoDecavswDeca}-\ref{figSDLScmpTLTnoDecavswDeca}. 

\begin{figure}[t]
\includegraphics[width=0.8\textwidth]{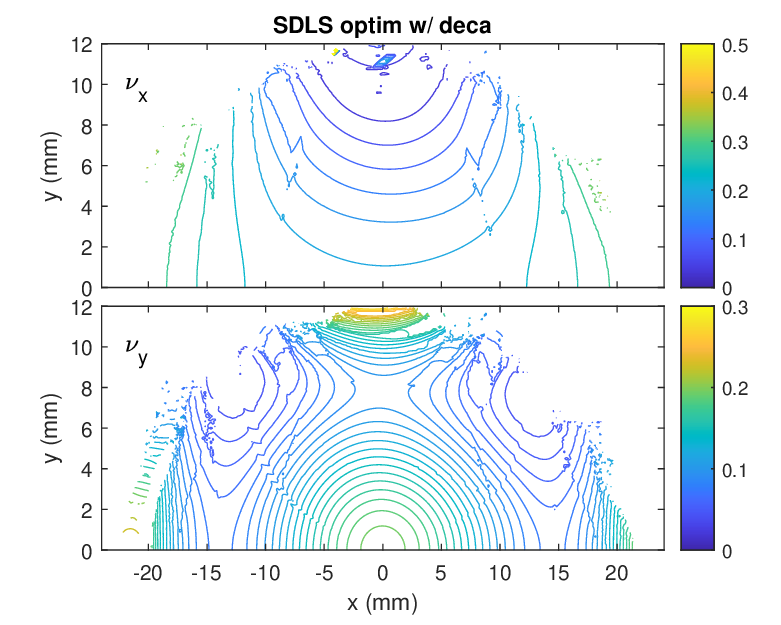}
\caption{Contour plots of horizontal (top) and vertical (bottom) betatron tunes for the optimized 
solution for SDLS. 
\label{figSDLSNuxyvsXYoptim46}}
\end{figure}

\begin{figure}[t]
\includegraphics[width=0.8\textwidth]{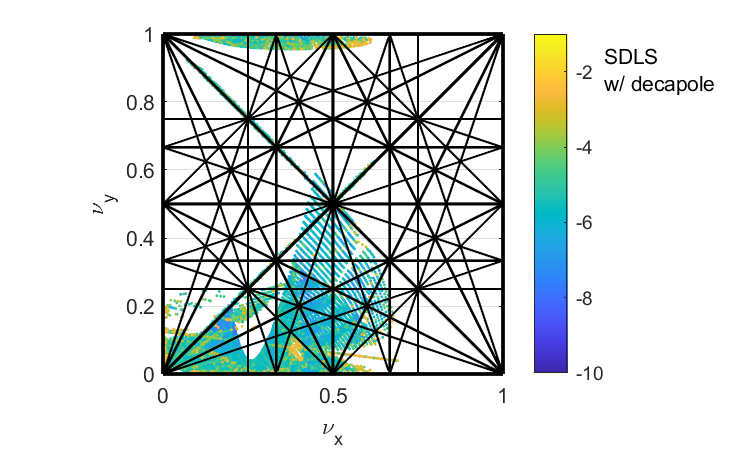}
\includegraphics[width=0.8\textwidth]{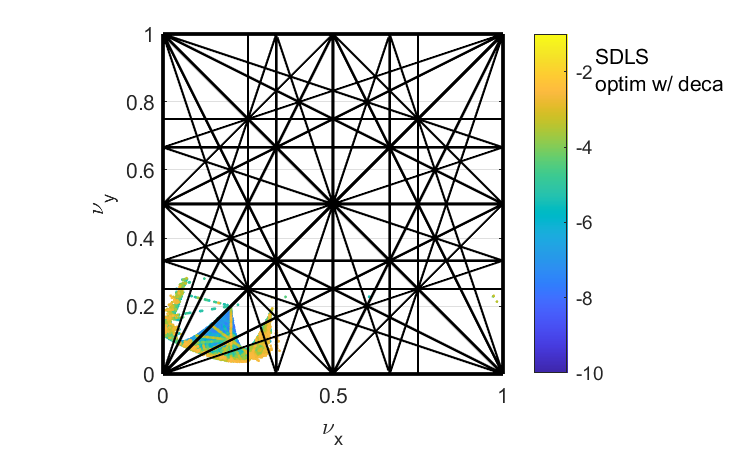}
\caption{The tune diagrams for the initial lattice (top) and the selected optimal solution (bottom)  for SDLS. 
The color code represents the tune diffusion. 
\label{figSDLSTuneDiagoptim46}}
\end{figure}

From Figure~\ref{figSDLScmpDAnoDecavswDeca}, it can be seen that compared to the original lattice (w/ decapole), the DA has a small gain in the negative 
$x$ side and a larger gain the vertical direction. 
Frequency map analysis of the selected optimal solution shows that it has a much smaller tune footprint. 
Figure~\ref{figSDLSNuxyvsXYoptim46} shows the betatron tunes vs. ($x$, $y$) launching positions, which can 
be compared to the initial solution in the bottom plots of Figure~\ref{figSDLScmpTuneXYSDLS}. 
The tune footprints in the tune diagram for the original lattice and the selected optimal solution 
are compared in Figure~\ref{figSDLSTuneDiagoptim46}. 
These figures show that not only the vertical tune of the optimal solution does not cross the integer below,
but also the horizontal tune is now confined to a much smaller range. 


\begin{figure}[t]
\includegraphics[width=0.8\textwidth]{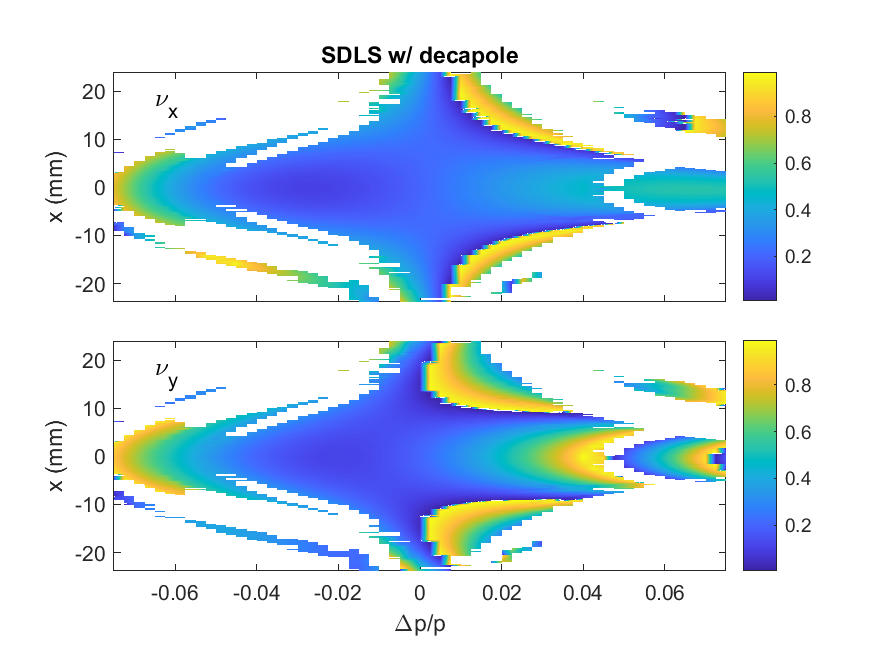}
\includegraphics[width=0.8\textwidth]{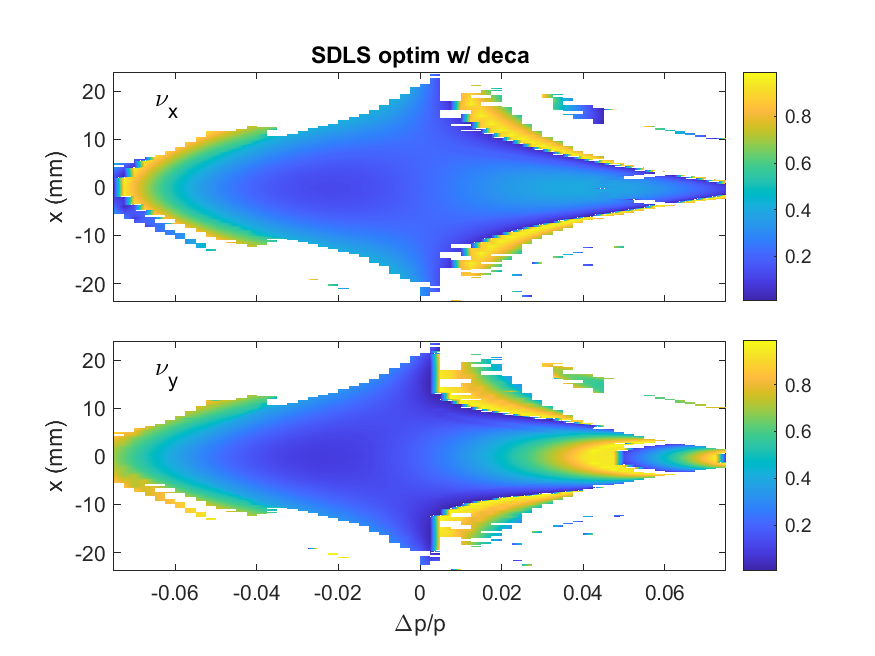}
\caption{Betatron tunes ($\nu_x$ and $\nu_y$) vs. ($x$, $\delta = \frac{\Delta p}{p}$) for the initial SDLS lattice (w/ decapole, in top plots) 
and the selected optimal solution (bottom plots). 
\label{figSDLSTunesVSxdpp46}}
\end{figure}

Figure \ref{figSDLScmpMAnoDecavswDeca} shows that the LMA is significantly enlarged throughout the cell in both the positive and negative sides. 
Consequently, the Touschek lifetime, evaluated for the same condition as in subsection~\ref{secSDLS:nl}, is increased. 
To  predict Touschek lifetime for realistic beam conditions, we include emittance growth due to 
IBS and with bunch lengthening by a factor of 4 through the use of a harmonic cavity. 
Full coupling is assumed and the emittances are $\epsilon_x=\epsilon_y=9.6$~pm for a bunch current of 
$I_b=0.1$~mA (see section~\ref{secBrightness}) while  
the bunch length is $\sigma_z=6.4$~mm. 
The Touschek lifetime (mean value of 25 seeds) is increased from 22.0 hrs to 94.9 hrs with nonlinear lattice optimization. 
In Figure~\ref{figSDLScmpTunesVsDpp}, the betatron tune dependence on momentum deviation of the optimized 
solution is compared to the initial lattice. 
It can be seen that the vertical tune of the optimized solution crosses the integer above at a larger 
momentum deviation (on the positive side) due to its reduced second order vertical chromaticity (from 
the initial value 226 to 202). This could in part explain the increase of LMA on the positive side. 
A closer examination of the betatron tune distribution in the ($x$, $\delta$) space 
(see Figure~\ref{figSDLSTunesVSxdpp46}) 
shows that 
the optimal solution also has larger stable regions for off-momentum particles, especially 
for $\delta<0$.

% >> [px0] = polyfit(a(5).XData(n), a(5).YData(n),3)  %SDLS w/ decapole
% px0 =   1.0e+03 *
%    -1.4230    0.0584    0.0063    0.0002
% >> [py0] = polyfit(a(4).XData(n), a(4).YData(n),3)  %SDLS w/ decapole
% py0 =  -23.7660  226.0481    9.1108    0.2006

% >> [px1] = polyfit(a(2).XData(n), a(2).YData(n),3)  %selected optimal
% px1 =   1.0e+03 *
%    -2.2788    0.0654    0.0062    0.0002
% py1] = polyfit(a(1).XData(n), a(1).YData(n),3)
% py1 = -117.3596  201.8634    9.6885    0.2001

\begin{figure}[t]
\includegraphics[width=0.8\textwidth]{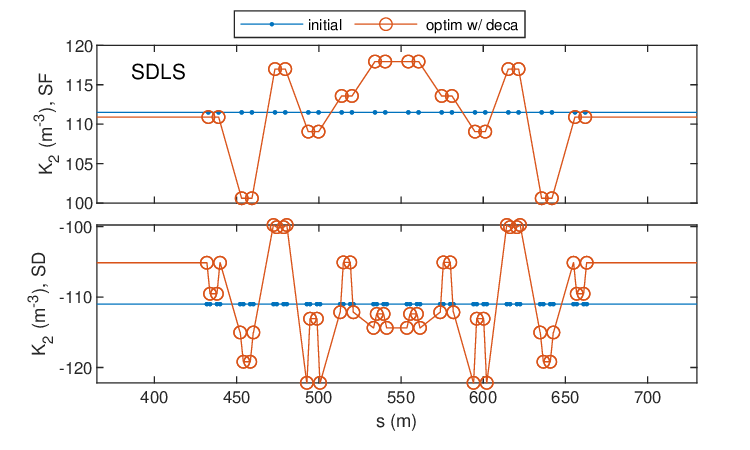}
\caption{The sextupole strengths for one arc of the SDLS lattice for the selected 
optimal solution and the initial lattice. 
\label{figSDLSK2pattenOptimReeval46}}
\end{figure}

While the initial lattice has equal strengths for sextupoles of the same family in 
all cells, the sextupole strengths for the optimized solution vary, as shown in 
Figure~\ref{figSDLSK2pattenOptimReeval46}. The focusing sextupole with the maximum strength 
is 6\% stronger than the initial strength, 
while the maximum strength of  defocusing sextupoles  increases by 10\%. 
The octupole and decapole strengths are varied by less than 1\%. 

\section{Beam brightness performance \label{secBrightness}}
The primary goal of a lattice upgrade for SSRL is to substantially boost the photon beam brightness. 
The photon beam brightness  depends on the electron beam emittances, optics functions at photon source points, photon beam energy and photon source properties, and the total beam current. 
The optics functions at the IDs are at the optimal values by design requirements for the three lattices. 
While the natural emittance is a property of the lattice, the actual beam emittances in the two transverse planes depend on the 
linear coupling one chooses to operate the machine with. 
In addition, IBS can increase the beam emittances when the bunch current is high. 
As IBS depends on linear coupling emittance ratio ($\epsilon_y/\epsilon_x$) and the longitudinal beam distribution, it is necessary to specify the 
relevant operating conditions when evaluating the photon beam brightness performance. 

We use the high-energy model in Ref.~\cite{BaneIBS} to calculate the IBS effects. 
For SSRLUP, a total beam current of 500~mA distributed evenly in 280 bunches is assumed.
This  operation mode  is the same as the current SPEAR3 storage ring. 
A linear coupling emittance ratio of $3.3\%$ is assumed, resulting in a vertical emittance of 10~pm at 
the low current limit. 
With the same 476.3~MHz rf system and a total rf voltage of 3~MV, the rms bunch length at low current is found to be 3.5~mm 
(with DW). With these conditions, the emittance growth by IBS for a 1.8~mA bunch current is calculated to be  $11\%$. 

\begin{figure}[t]
\includegraphics[width=0.8\textwidth]{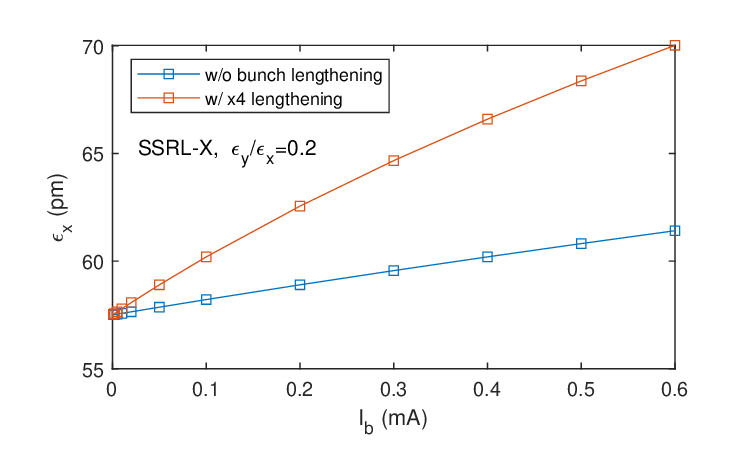}
\caption{The increase of emittances due to IBS for SSRL-X (with total damping wiggler length of 20 m), with coupling ratio 
$\epsilon_y/\epsilon_x=20\%$. Two cases, one without bunch lengthening (blue), 
the other with bunch lengthening by a factor of 4 (red), are shown. 
\label{figSSRLXIBSvsIb}}
\end{figure}

For SSRL-X, a 476 MHz rf system with a total rf voltage of 3~MV is assumed. 
A $20\%$ coupling ratio is assumed for this lattice to maximize photon beam brightness. 
Figure~\ref{figSSRLXIBSvsIb} shows the increase of horizontal emittance 
as functions of the bunch current, with or without bunch lengthening by a harmonic cavity. Damping wiggler of a total length of 20~m is assumed. 
If the ring operates with 700 (out of 932 buckets) uniformly filled bunches, 
the emittance would be 61.2~pm for a total current of 400~mA (w/ bunch current 0.57~mA) with the bunch lengthened 
 to 9~mm.  

\begin{figure}[t]
\includegraphics[width=0.8\textwidth]{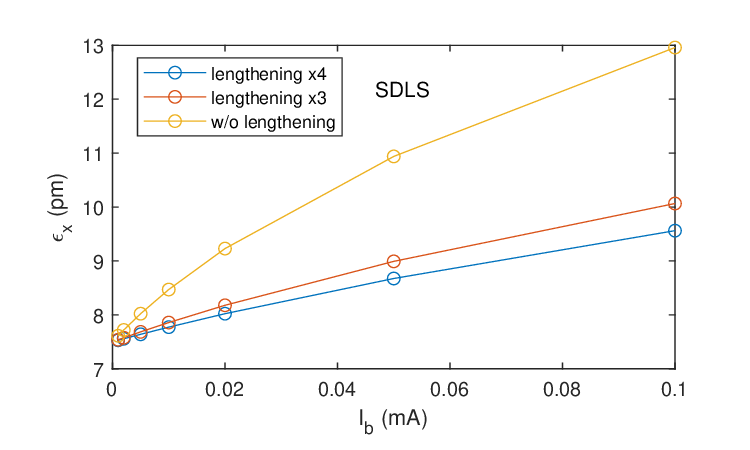}
\caption{The increase of emittances ($\epsilon_x=\epsilon_y$ with 100\% coupling) by IBS for SDLS (with total damping wiggler length of 70 m). 
Three cases, without bunch lengthening, or with bunch lengthened by a factor of 3 or 4, are compared. 
\label{figSDLSIBSvsIb}}
\end{figure}

The IBS effects on SDLS are found to be larger. 
We assume a 476~MHz rf system with an rf voltage of 6~MV. The rms bunch length would be 1.3~mm without bunch lengthening. 
Figure~\ref{figSDLSIBSvsIb} shows the increase of emittance with bunch current for the case with damping wigglers and full coupling. 
At the low current limit, the
emittances for both transverse planes would be 7.5~pm. 
However, if there is no bunch lengthening by a harmonic cavity, the IBS would cause the emittance to grow by 75\% for a bunch current of 0.1~mA. 
Therefore, it would be necessary to employ a bunch lengthening harmonic cavity to lengthen the bunch. 
Figure~\ref{figSDLSIBSvsIb} also shows the emittance increase for the two cases when the bunch is lengthened by a factor of 
3 or 4. 
For bunch lengthening by a factor of 4 and a bunch current of 0.1~mA, 
the emittance becomes 9.6~pm and the momentum spread increases by  3.4\% (to $\sigma_\delta=9.1\times10^{-3}$). 
%With a bunch current of 0.1~mA, a total current of 300~mA can be stored in 3000 bunches (out of 3479 buckets). 
The emittances are below the diffraction limited level for the photon energy of 10~keV.

\begin{figure}[t]
\includegraphics[width=0.8\textwidth]{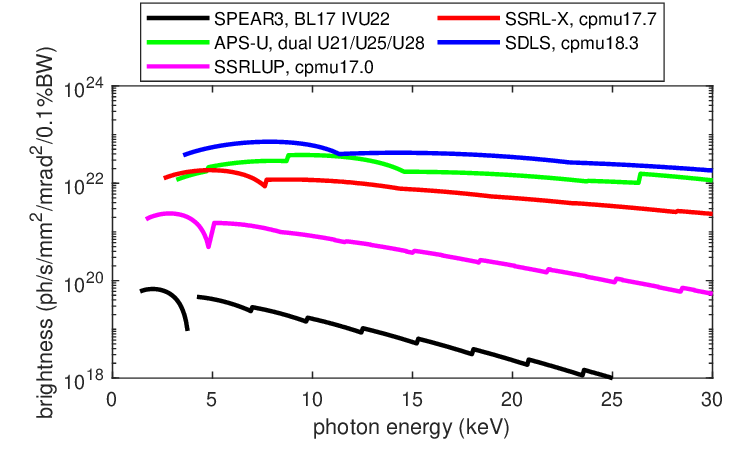}
\caption{Comparison of photon beam brightness of the upgrade lattice options to 
the existing SPEAR3 and the future APS-U lattice. 
Insertion device parameters vary as appropriate to reflect the reality (SPEAR3), the current plan (APS-U), or 
a reasonable prediction suitable for the corresponding lattices (SSRL upgrade). 
\label{figCmpBrightness}}
\end{figure}

With the IBS calculation, we obtain the beam parameters for the corresponding total beam current, which can be used in photon beam 
brightness calculations. 
For each upgrade lattice option, an undulator with reasonable parameters is assumed to best exploit the beam 
condition. 
For SSRLUP, a 2.1~m device with 126 periods and a peak field of 1.26~T (w/ undulator parameter $K=2.00$) is assumed. 
For SSRL-X, a 3.1~m device with 173 periods and a peak field of 1.31~T (w/ $K=2.17$) is assumed. 
For SDLS, a 3.6~m device with 194 period and a peak field of 1.35~T (w/ $K=2.31$) is assumed. 
In Figure~\ref{figCmpBrightness}, the photon beam brightness curves of the upgrade lattices are compared to that of the 
current SPEAR3 and the future APS upgrade~\cite{APSU41pm} lattices. 

\section{Conclusion}
In this study, three lattice options were designed for a future upgrade of the Stanford 
Synchrotron Radiation Lightsource. 
The three options can be used to accommodate different funding scenarios as they 
differ in the ring size, siting option, and cost. The beam energies also differ as needed to take 
advantage of the ring size. 
The lattices are based on the recently proposed hybrid six-bend achromat (H6BA) cell structure~\cite{Pantaleo2023PRAB}, which enables simultaneously reaching low emittances and 
excellent nonlinear dynamics performances. 
Key lattice parameters are listed in Table~\ref{tab:RingPara}.

\begin{table*}[htbp] 
\begin{minipage}{\textwidth}
\caption{Ring parameters for the three lattice options} 
\label{tab:RingPara} 
\centering
%\begin{center} 
\begin{tabular}{l|c|c|c} 
\hline 
Parameter &  SSRLUP & SSRL-X  & SDLS \\
& & & \\
\hline
Circumference (m) & 234.0 & 586.7  &2189.8 \\
Beam energy (GeV) & 3 & 4 & 5 \\
Momentum compaction, ($\alpha_c$, $10^{-4}$) & 4.44 & 1.09 & 0.26 \\
Horizontal tune, $\nu_x$ &32.25  & 78.2 & 133.2 \\
Vertical tune, $\nu_y$ &15.17 & 37.2 & 71.2 \\
Horizontal beta function at ID, $\beta_x$ (m) &2.9 & 2.0 & 2.0 \\
Vertical function at ID, $\beta_y$ (m) &1.8 & 2.0 & 2.0 \\
Horizontal beta function at injection, $\beta_x$ (m) &11.2 & 20.0 & 45.5 \\
Vertical function at injection, $\beta_y$ (m) &3.6 & 3.0 & 17.0 \\
%Betatron tunes & & & \\
%\hspace{3mm} $\nu_x$ &32.25  & & \\
%\hspace{3mm} $\nu_y$ &15.167 & & \\
Bare lattice: & & & \\
\hspace{3mm} Emittance (pm) & 364 & 86 & 28 \\
\hspace{3mm} Momentum spread $\sigma_\delta$ ($\times10^{-3}$) &  1.07 & 0.67 & 0.63 \\
\hspace{3mm} Energy loss $U_0$ (MeV) &  0.64 & 0.59 & 0.68 \\
\hspace{3mm} Damping partition\footnote{For all three lattices, $J_y=1$ and $J_z=3-J_x$. } $J_x$ &  1.859 & 1.414 & 1.675 \\
w/ damping wigglers\footnote{For SSRLUP, the existing wigglers in SPEAR3 are assumed. For SSRL-X and SDLS,  damping wigglers 
with peak field of 1~T and wiggler period of 12.4~cm are assumed, with total DW length of 20 m and 70 m for SSRLX and SDLS, respectively. 
}: & & & \\
\hspace{3mm} Emittance (pm) & 310 & 69.2 & 15.0 \\
\hspace{3mm} Momentum spread $\sigma_\delta$ ($\times10^{-3}$) & 1.04 & 0.73  & 0.88 \\
\hspace{3mm} Energy loss $U_0$ (MeV) &   0.87 & 0.79  & 1.79 \\
\hspace{3mm} Damping partition $J_x$ &  1.629 & 1.308 & 1.257 \\
\hline
RF\footnote{RF frequency is assumed to be 476.3 MHz for all three cases.} 
    voltage (MV) & 3 & 3 & 6 \\
Harmonic number & 372 & 932 & 3479 \\
Bunch length (low current), $\sigma_{z0}$ (mm) & 3.5  & 2.2  & 1.3 \\ 
Synchrotron tune, $\nu_s$ & 0.0051  & 0.0035 & 0.0042  \\
Total current (mA) & 500 & 400 & 300 \\
Number of bunches & 280 & 700 &  3000 \\
Bunch current, $I_b$ (mA) & 1.79 & 0.57  & 0.10 \\
Coupling ratio, $\frac{\epsilon_y}{\epsilon_x}$ & 0.033 & 0.2  & 1.0 \\
w/ IBS & & & \\
\hspace{3mm} Bunch length\footnote{With bunch lengthening by a factor of $\times4$ for SSRLX and SDLS.}, $\sigma_{z}$ (mm) & 3.6  & 9.0  & 6.4 \\ 
\hspace{3mm} Horizontal emittance, $\epsilon_x$ (pm) & 333.7 & 61.2 & 9.6 \\
\hspace{3mm} Vertical emittance, $\epsilon_y$ (pm) & 11.0 & 12.2 & 9.6 \\
\hspace{3mm} Momentum spread $\sigma_\delta$ ($\times10^{-3}$) & 1.07 & 0.74  & 0.91 \\
\hspace{3mm} Touschek lifetime (hr) & 4.9 & 44.7 & 94.9 \\
\hline 

\end{tabular} 

%\end{center} 
\end{minipage}
\end{table*} 

High beta injection straights are employed to enable off-axis injection. 
By requiring transparent matching conditions, the lattices maintain excellent nonlinear 
dynamics performance despite the loss of linear optics periodicity. 
The dynamic aperture (16~mm) and Touschek lifetime (4.9~hr for a 500-mA total beam current with 3.3\% coupling) of the SSRLUP lattice are similar to that of 
the current SPEAR3.
Numerical optimization of nonlinear magnets were carried out to improve the dynamic aperture 
and Touschek lifetime for SSRL-X and SDLS. 
For the SSRL-X lattice, numerical optimization eliminates the need of decapoles and the optimized 
solution achieves a better Touschek lifetime. 
For the SDLS lattice,  octupole and decapole magnets play a significant role in shaping the tune footprint 
of the beam. Numerical optimization leads to substantial improvement in Touschek lifetime. 
The Touschek lifetime reaches 44.7~hr (SSRL-X, with 400~mA in 700 bunches ) and 94.9~hr (SDLS, with 300~mA in 3000 bunches) for the two ultra low emittance rings, respectively, with the use of bunch lengthening harmonic cavities. 
All three latices have large dynamic apertures which can easily support traditional off-axis injection. 
The large dynamic aperture and long beam lifetime are useful for possible advanced beam manipulation schemes, 
such as 2-frequency crab cavity for short pulses~\cite{ZHOLENTS2015111,Huang2FCC2019}. 
The nonlinear dynamics performances of the lattices are evaluated with random linear errors (25 seeds) 
added to lattice. Multipole errors are to be included in future studies. 

Intra-beam scattering (IBS) has only a small impact to the SSRLUP beam conditions. However, 
IBS would increase the emittances by  19\% for an SSRL-X bunch of 0.57-mA with a coupling ratio of 20\% if not using bunch lengthening harmonic cavity. With bunch lengthening by a factor of 4, the IBS emittance growth is reduced to 6\%. 
For SDLS, bunch lengthening is also necessary to avoid emittance blowup by IBS. 
Calculation shows that even with bunch lengthening by a factor of 4, the emittance for a 0.1-mA bunch 
still increases from 7.5 to 9.6~pm in both planes (with full coupling). 

\section*{Acknowledgments}
%\begin{acknowledgments}
P.R. designed all three lattices. 
J.S. and T.R. provided input on siting constraints and beamline and user science preferences and requirements. 
X.H. and J.K. evaluated lattice performances as input during design iterations. 
T.R. calculated the photon brightness curves. 
X.H. performed nonlinear lattice optimization and evaluation, calculated IBS effects, and drafted the paper. 

  This work was supported by the U.S. Department of Energy, Office of
  Science, Office of Basic Energy Sciences, under Contract No.
  DE-AC02-76SF00515.  
%\end{acknowledgments}

\clearpage
%\nocite{*}
\bibliography{dama_ref}% Produces the bibliography via BibTeX.

\begin{thebibliography}{10}
\expandafter\ifx\csname url\endcsname\relax
  \def\url#1{\texttt{#1}}\fi
\expandafter\ifx\csname urlprefix\endcsname\relax\def\urlprefix{URL }\fi
\expandafter\ifx\csname href\endcsname\relax
  \def\href#1#2{#2} \def\path#1{#1}\fi

\bibitem{HettelEPAC04}
R.~Hettel, et~al., The completion of {SPEAR3}, in: Proceedings of EPAC'04,
  Lucerne, Switzerland, 2004, pp. 2451--2453.

\bibitem{EinfeldPAC95}
D.~Einfeld, J.~Schaper, M.~Plesko, Design of a diffraction limited light source
  {(DIFL)}, in: Proceedings of PAC'95, Dallas, TX, 1995, pp. 177--179.

\bibitem{LeemanMAXIV}
S.~C. Leemann, A.~Andersson, M.~Eriksson, L.-J. Lindgren, E.~Wall\'en,
  J.~Bengtsson, A.~Streun,
  \href{https://link.aps.org/doi/10.1103/PhysRevSTAB.12.120701}{Beam dynamics
  and expected performance of {Sweden's} new storage-ring light source: {MAX
  IV}}, Phys. Rev. ST Accel. Beams 12 (2009) 120701.
\newblock \href {https://doi.org/10.1103/PhysRevSTAB.12.120701}
  {\path{doi:10.1103/PhysRevSTAB.12.120701}}.
\newline\urlprefix\url{https://link.aps.org/doi/10.1103/PhysRevSTAB.12.120701}

\bibitem{ESRFHMBA}
L.~Farvacque, N.~Carmignani, J.~Chavanne, A.~Franchi, G.~L. Bec, S.~Liuzzo,
  B.~Nash, T.~P.~P. Raimondi, A low-emittance lattice for the {ESRF}, in:
  Proceedings of IPAC 2013, Shanghai, China, 2013, pp. 79--81.

\bibitem{APSU41pm}
M.~Borland, Y.~Sun, V.~Sajaev, R.~R. Lindberg, T.~Berenc, Lower emittance
  lattice for the {Aadvanced Photon Source} upgrade using reverse bending
  magnets, in: Proceedings of NAPAC 2016, Chicago, IL, USA, 2016, pp. 877--880.

\bibitem{ALSU}
C.~Steier, et~al., Status of the conceptual design of {ALS-U}, in: Proceedings
  of IPAC 2017, Copenhagen, Denmark, 2017, pp. 2824--2826.

\bibitem{HEPS}
Y.~Jiao, G.~Xu, Y.~Peng, S.~Chen, J.~Li, Q.~Qin, J.~Wang, C.~Y, Evolution of
  the lattice design for the {High Energy Photon Source}, in: Proceedings of
  IPAC 2018, Vancouver, BC, Canad, 2018, pp. 1363--1366.

\bibitem{PETRAIV}
I.~Agapov, S.~Antipov, R.~Bartolini, R.~Brinkmann, Y.-C. Chae, D.~Einfeld,
  T.~Hellert, M.~Hüening, M.~Jebramcik, J.~Keil, C.~Li, R.~Wanzenberg, {PETRA
  IV} storage ring design, in: Proceedings of IPAC 2022, Bangkok, Thailand,
  2022, pp. 1431--1434.

\bibitem{ELETTRA2}
E.~Karantzoulis, The diffraction limited light source {ELETTRA2} 2.0, in:
  Proceedings of IPAC 2017, Copenhagen, Denmark, 2017, pp. 2660--2663.

\bibitem{DIAMOND2}
H.~Ghasem, I.~P.~S. Martin, B.~Singh, Progress with the {DIAMOND-II} storage
  ring lattice, in: Proceedings of IPAC 2021, Campinas, SP, Brazil, 2021, pp.
  3973--3976.

\bibitem{SOLEILU2}
A.~Nadji, L.~S. Nadolski,
  \href{https://doi.org/10.1080/08940886.2023.2186661}{Upgrade project of the
  {SOLEIL} accelerator complex}, Synchrotron Radiation News 36~(1) (2023)
  10--15.
\newblock \href
  {http://arxiv.org/abs/https://doi.org/10.1080/08940886.2023.2186661}
  {\path{arXiv:https://doi.org/10.1080/08940886.2023.2186661}}, \href
  {https://doi.org/10.1080/08940886.2023.2186661}
  {\path{doi:10.1080/08940886.2023.2186661}}.
\newline\urlprefix\url{https://doi.org/10.1080/08940886.2023.2186661}

\bibitem{Pantaleo2023PRAB}
P.~Raimondi, S.~M. Liuzzo,
  \href{https://link.aps.org/doi/10.1103/PhysRevAccelBeams.26.021601}{Toward a
  diffraction limited light source}, Phys. Rev. Accel. Beams 26 (2023) 021601.
\newblock \href {https://doi.org/10.1103/PhysRevAccelBeams.26.021601}
  {\path{doi:10.1103/PhysRevAccelBeams.26.021601}}.
\newline\urlprefix\url{https://link.aps.org/doi/10.1103/PhysRevAccelBeams.26.021601}

\bibitem{ESRFEBS}
J.~Biasci, J.~Bouteille, N.~Carmignani, J.~Chavanne, D.~Coulon, Y.~Dabin,
  F.~Ewald, L.~Farvacque, L.~Goirand, M.~Hahn, J.~Jacob, G.~LeBec, S.~Liuzzo,
  B.~Nash, H.~Pedroso-Marques, T.~Perron, E.~Plouviez, P.~Raimondi, J.~Revol,
  K.~Scheidt, V.~Serrière, A low-emittance lattice for the {ESRF}, Synchrotron
  Radiation News 27~(6) (2014) 8--12.
\newblock \href {https://doi.org/10.1080/08940886.2014.970931}
  {\path{doi:10.1080/08940886.2014.970931}}.

\bibitem{EBScomm}
P.~Raimondi, N.~Carmignani, L.~R. Carver, J.~Chavanne, L.~Farvacque, G.~Le~Bec,
  D.~Martin, S.~M. Liuzzo, T.~Perron, S.~White,
  \href{https://link.aps.org/doi/10.1103/PhysRevAccelBeams.24.110701}{Commissioning
  of the hybrid multibend achromat lattice at the {European Synchrotron
  Radiation Facility}}, Phys. Rev. Accel. Beams 24 (2021) 110701.
\newblock \href {https://doi.org/10.1103/PhysRevAccelBeams.24.110701}
  {\path{doi:10.1103/PhysRevAccelBeams.24.110701}}.
\newline\urlprefix\url{https://link.aps.org/doi/10.1103/PhysRevAccelBeams.24.110701}

\bibitem{yang2008global}
L.~Yang, D.~Robin, F.~Sannibale, C.~Steier, W.~Wan, Global optimization of the
  magnetic lattice using genetic algorithms, in: EPAC, Vol.~8, 2008, p. 3050.

\bibitem{borland2010direct}
M.~Borland, V.~Sajaev, L.~Emery, A.~Xiao, Direct methods of optimization of
  storage ring dynamic and momentum aperture, in: PAC09, 2009, pp. 3850--3852.

\bibitem{HUANG201448}
X.~Huang, J.~Safranek,
  \href{http://www.sciencedirect.com/science/article/pii/S0168900214004914}{Nonlinear
  dynamics optimization with particle swarm and genetic algorithms for {SPEAR3}
  emittance upgrade}, Nuclear Instruments and Methods in Physics Research
  Section A: Accelerators, Spectrometers, Detectors and Associated Equipment
  757 (2014) 48 -- 53.
\newblock \href {https://doi.org/https://doi.org/10.1016/j.nima.2014.04.078}
  {\path{doi:https://doi.org/10.1016/j.nima.2014.04.078}}.
\newline\urlprefix\url{http://www.sciencedirect.com/science/article/pii/S0168900214004914}

\bibitem{TerebiloAT01}
A.~Terebilo, Accelerator modeling with {Matlab Accelerator Toolbox}, in:
  Proceedings of PAC2001, Chicago, IL, 2001, pp. 3203--3205.

\bibitem{LASKAR1993257naff}
J.~Laskar,
  \href{http://www.sciencedirect.com/science/article/pii/016727899390210R}{Frequency
  analysis for multi-dimensional systems. {Global} dynamics and diffusion},
  Physica D: Nonlinear Phenomena 67~(1) (1993) 257 -- 281.
\newblock \href {https://doi.org/https://doi.org/10.1016/0167-2789(93)90210-R}
  {\path{doi:https://doi.org/10.1016/0167-2789(93)90210-R}}.
\newline\urlprefix\url{http://www.sciencedirect.com/science/article/pii/016727899390210R}

\bibitem{FMALaska}
J.~Laskar, Frequency map analysis and particle accelerators, in: Proceedings of
  PAC'03, 2003, pp. 378--382.

\bibitem{Montague}
B.~W. Montague, Linear optics for improved chromaticity correction, Tech. rep.
  (1979).

\bibitem{hettel2008ideas}
R.~Hettel, K.~Bane, L.~Bentson, K.~Bertsche, S.~Brennan, Y.~Cai, A.~Chao,
  S.~DeBarger, V.~Dolgashev, X.~Huang, et~al., Ideas for a future {PEP-X} light
  source, in: Proceedings of EPAC'08, Genoa, Italy, 2008, pp. 2031--2033.

\bibitem{HettelPAC09}
R.~Hettel, et~al., Concept for the {PEP-X} light source, in: Proceedings of
  PAC'09, Vancouver, BC, Canada, 2009, pp. 2297--2299.

\bibitem{WANG201130}
M.-H. Wang, Y.~Nosochkov, K.~Bane, Y.~Cai, R.~Hettel, X.~Huang,
  \href{https://www.sciencedirect.com/science/article/pii/S0168900211000556}{Lattice
  design and optimization for the {PEP-X} ultra low emittance storage ring at
  {SLAC}}, Nuclear Instruments and Methods in Physics Research Section A:
  Accelerators, Spectrometers, Detectors and Associated Equipment 649~(1)
  (2011) 30--34, national Synchrotron Radiation Instrumentation conference in
  2010.
\newblock \href {https://doi.org/https://doi.org/10.1016/j.nima.2011.01.011}
  {\path{doi:https://doi.org/10.1016/j.nima.2011.01.011}}.
\newline\urlprefix\url{https://www.sciencedirect.com/science/article/pii/S0168900211000556}

\bibitem{cai2012ultimate}
Y.~Cai, K.~Bane, R.~Hettel, Y.~Nosochkov, M.-H. Wang, M.~Borland, Ultimate
  storage ring based on fourth-order geometric achromats, Physical Review
  Special Topics-Accelerators and Beams 15~(5) (2012) 054002.

\bibitem{HuangDA2015}
X.~Huang, J.~Safranek,
  \href{https://link.aps.org/doi/10.1103/PhysRevSTAB.18.084001}{Online
  optimization of storage ring nonlinear beam dynamics}, Phys. Rev. ST Accel.
  Beams 18 (2015) 084001.
\newblock \href {https://doi.org/10.1103/PhysRevSTAB.18.084001}
  {\path{doi:10.1103/PhysRevSTAB.18.084001}}.
\newline\urlprefix\url{https://link.aps.org/doi/10.1103/PhysRevSTAB.18.084001}

\bibitem{BaneIBS}
K.~Bane, A simplified model of intrabeam scattering, in: Proceedings of {EPAC}
  2002, Paris, France, IEEE, 2002, pp. 1443--1445.

\bibitem{APSU}
{Advanced Photon Source Upgrade Project Preliminary Design Report, Argonne
  National Laboratory}, {APSU-2.01-RPT-002} (September 2017).

\bibitem{ZHOLENTS2015111}
A.~Zholents,
  \href{https://www.sciencedirect.com/science/article/pii/S0168900215008451}{A
  new possibility for production of sub-picosecond x-ray pulses using a time
  dependent radio frequency orbit deflection}, Nuclear Instruments and Methods
  in Physics Research Section A: Accelerators, Spectrometers, Detectors and
  Associated Equipment 798 (2015) 111--116.
\newblock \href {https://doi.org/https://doi.org/10.1016/j.nima.2015.07.016}
  {\path{doi:https://doi.org/10.1016/j.nima.2015.07.016}}.
\newline\urlprefix\url{https://www.sciencedirect.com/science/article/pii/S0168900215008451}

\bibitem{Huang2FCC2019}
X.~Huang, B.~Hettel, T.~Rabedeau, J.~Safranek, J.~Sebek, K.~Tian, K.~P.
  Wootton, A.~Zholents,
  \href{https://link.aps.org/doi/10.1103/PhysRevAccelBeams.22.090703}{Beam
  dynamics issues for the two-frequency crab cavity short pulse scheme}, Phys.
  Rev. Accel. Beams 22 (2019) 090703.
\newblock \href {https://doi.org/10.1103/PhysRevAccelBeams.22.090703}
  {\path{doi:10.1103/PhysRevAccelBeams.22.090703}}.
\newline\urlprefix\url{https://link.aps.org/doi/10.1103/PhysRevAccelBeams.22.090703}

\end{thebibliography}

\end{document}